\documentclass[fleqn,usenatbib]{mnras}
\usepackage[utf8]{inputenc}
\usepackage{mathrsfs}
\usepackage{natbib}
\usepackage{amsmath,amssymb}
\usepackage{amsfonts}
\usepackage{graphicx}
\usepackage{hyperref}
\usepackage[usenames,dvipsnames]{color}
\usepackage{mathrsfs}

\makeatletter

\usepackage{mathrsfs}

\newcommand{\abs}[1]{\left\vert#1\right\vert}
\newcommand{\set}[1]{\left\{#1\right\}}

\newcommand{\eps}{\varepsilon}

\newcommand{\arccosh}{\mathrm{arccosh}}
\newcommand{\zhat}{\mathbf{\hat{z}}}

\newcommand{\bin}{\textrm{bin}}

\newcommand{\scri}{\mathscr{I}}

\makeatother

\title[3-Body Binary Formation]{Three-Body Binary Formation in Clusters: Analytical Theory}

\author[Y. B. Ginat and H. B. Perets]{Yonadav Barry Ginat,$^{1,2 }$\thanks{E-mail: yb.ginat@physics.ox.ac.uk}
and
Hagai B. Perets$^{3}$
\\
$^{1}$Rudolf Peierls Centre for Theoretical Physics, University of Oxford, Parks Road, Oxford, OX1 3PU, United Kingdom \\ 
$^{2}$New College, Holywell Street, Oxford, OX1 3BN, United Kingdom \\
$^{3}$Faculty of Physics, Technion -- Israel Institute of Technology,
Haifa, 3200003, Israel
}

\date{Accepted XXX. Received YYY; in original form ZZZ}

\pubyear{2024}
\begin{document}
\label{firstpage}
\pagerange{\pageref{firstpage}--\pageref{lastpage}}
\maketitle

\begin{abstract}
Binary formation in clusters through triple encounters between three unbound stars, `three-body' binary formation, is one of the main dynamical formation processes of binary systems in dense environments. In this paper, we use an analytical probabilistic approach to study the process for the equal mass case and calculate a probability distribution for the orbital parameters of three-body-formed binaries, as well as their formation rate. For the first time, we give closed-form analytical expressions to the full orbital parameter distribution, accounting for both energy and angular momentum conservation. This calculation relies on the sensitive dependence of the outcomes of three-body scatterings on the initial conditions: here we compute the rate of three-body binaries from ergodic interactions, which allow for an analytical derivation of the distribution of orbital parameters of the binaries thus created. We find that soft binaries are highly favoured in this process and that these binaries have a super-thermal eccentricity distribution, while the few hard three-body binaries have an eccentricity distribution much closer to thermal. The analytical results predict and reproduce simulation results of three-body scattering experiments in the literature well.  
\end{abstract}

\begin{keywords}
galaxies: clusters: general -- chaos -- (stars:) binaries: general -- stars: kinematics and dynamics
\end{keywords}



\section{Introduction}

Binary stars are among the most important astrophysical systems, playing a key part in stellar evolution, and in the dynamics of dense stellar systems. They have a major role in the evolution of globular clusters \citep[e.g.][]{HeggieHut2003,Binney}, and effectively serve to stop core-collapse \citep[e.g.][]{Hut1996,GoodmanHut1989}. Among the leading channels for the formation of binaries in clusters are primordial binaries formed in the first epoch of star formation in the cluster \citep[e.g.][]{Shuetal1987},  tidal captures \citep[e.g.][]{Fabianetal1975,PressTeukolsky1977,Generozovetal2018}, and dynamical formation of binaries, via three-body interaction -- so-called `three-body' binaries \citep[e.g.][]{Mansbach1970,AarsethHeggie1976,Stodolkiewicz1986,GoodmanHut1993,ZwartMcMillan2000,HeggieHut2003,Atallahetal2024}, which are the topic of this paper -- see, e.g. \cite{Pooleyetal2003} for observational evidence for this process. In such an interaction, three unbound bodies exchange energy, such that at the end two of them remain bound, with the third one serving as a catalyst, an energy reservoir where excess energy is deposited. 
Incidentally, it has been proposed recently that if sufficient amounts of gas are present in a cluster, this could accelerate binary formation by serving as a dissipation mechanism -- an alternative energy reservoir \citep{Rozneretal2023}. 

In this paper, we wish to study three-body binary formation from an analytical perspective. We seek to offer a complimentary calculation to those of \cite{AarsethHeggie1976}, who approached the problem in an impulsive approximation, and \cite{GoodmanHut1993}, who used detailed balance to convert an analysis of the inverse problem -- that of the ionisation of a binary by a single star, to three-body binary formation. Here, instead, we would like to harness the chaotic nature of the three-body problem \citep[e.g.][]{ValtonenKarttunen2006} to derive probabilistic results on three-body binary formation. While previous analytical works on the probabilistic solution of the three-body problem \citep[e.g.][]{Monaghan1976a,Monaghan1976b,ValtonenKarttunen2006,StoneLeigh2019,GinatPerets2021a,GinatPerets2021b,Kol2021} have studied the negative energy case, i.e. a binary-single encounter, here we study the positive-energy case -- the formation of a binary from three initially unbound stars. This approach allows us to calculate not only the binary formation rate but also the joint eccentricity and semi-major axis distribution, in a way that accounts for angular momentum conservation as well as energy conservation, during the three-body interaction. 

The three-body binary formation can be automatically accounted for in $N$-body simulations of globular clusters, if close encounters are well resolved, and not smeared by artificial softening \citep[e.g.][]{vanAlbada1968,Aarseth1969,BreenHeggie2012a,BreenHeggie2012b,Tanikawa2013,Wangetal2016,Parketal2017,Kumamotoetal2019,Arcaseddaetal2023}, but not necessarily in Monte-Carlo simulations \citep[e.g.][]{Giersz1998,Ivanovaetal2005,Morscheretal2015,Gelleretal2019,Hongetal2020,Rodriguezetal2022,Weatherfordetal2023}, which often resort to implementing a prescription for three-body binary formation \citep{Ivanovaetal2005}. The former simulations are often limited, as modelling realistic globular clusters, requires expensive long-term simulations, as such clusters are old, highly populated, and diverse -- and thus too computationally expensive -- to be accurately modelled by direct $N$-body simulations. Moreover, few-body simulations do not provide an inherent understanding of the formation process and its dependence on the properties of the encounters.  It is thus important to derive an analytic understanding of the processes. Moreover, such analysis can test and replace semi-analytic prescriptions used in Monte-Carlo simulations, which are more versatile than direct $N$-body ones. 

The paper is structured as follows: we start by calculating the probability distribution of binary parameters, \emph{given} fixed values of the conserved quantities -- the energy and angular momentum (as well as the masses) in \S \ref{sec: f_bin for fixed kappa}. This allows us to study the properties of binaries formed this way as a function of these parameters. Then, in \S \ref{sec:rate}, we consider the binary formation rate in a system, e.g. a globular cluster, given a distribution of initial positions and velocities. In both sections, we endeavour to keep the lengthier calculations to appendices, while quoting the physical results in the main text. In \S \ref{sec:discussion} we summarise the results and highlight their implications. In this paper, we consider strictly the equal mass case, although many of the formul\ae derived here extend immediately to the unequal case (in particular, most of \S \ref{sec: f_bin for fixed kappa}). 

In the final stages of this work, \cite{Atallahetal2024} published a \emph{numerical} study of the three-body binary formation. They analysed the results of three-body numerical experiments with positive energy, which provides us with direct data to test our analytic predictions. We therefore structured the calculation of the binary-formation rate in \S \ref{sec:rate} to allow for a simple comparison with their results. As the results of \cite{Atallahetal2024} are integrated over a distribution of energies and angular momenta, the calculations of \S \ref{sec: f_bin for fixed kappa} are not immediately comparable with them, (but as tested via their integrals in \S \ref{sec:rate}). 

\section{Resonant Unbound Three-Body Scatterings}
\label{sec: f_bin for fixed kappa}
Consider three bodies coming from infinity, with masses $m_1$, $m_2$, and $m_3$, and total energy $E>0$ and angular momentum $\mathbf{J}$. For point particles, the possible scattering outcomes are: (i) three unbound stars, or (ii) a bound binary and a single unbound star. In this section, we focus on the probability for the outcome (ii), for a given set of conserved quantities, i.e. for given values of $E$, and the total angular momentum (in the centre-of-mass frame), $\mathbf{J}$. To do so, we have to describe the allowed region of phase-space (in \S\S \ref{subsec:Cut-Offs} and \ref{subsec: additional phase-space constraints}). The resultant orbital parameter distribution follows, from which we derive scaling relations with the spatial cut-off (see below) and the hard-soft-boundary. 

The phase-space volume for a positive-energy three-body system is infinite, so we must require that the three bodies interact inside a region of size $R_0$, which will be described in \S \ref{subsec:Cut-Offs} below. For now, suffice it to say that if they don't, then for $R_0 \to \infty$, the interaction is not chaotic, so indeed these should be excluded. 

If the outcome is a bound binary and a single star, then we label the remnant binary parameters with a sub-script $b$, and those of the orbit of the single star about the binary's centre of mass with a sub-script $s$. However, we still do that if there are three unbound stars, simply reserving the sub-script $b$ for the pair with the lowest two-body energy, and we still refer to this pair as a `binary', albeit unbound.

Then, in principle, as long as the outcome is highly sensitive to initial conditions, one can still apply the theory of approximate solutions to the three-body problem, similar to situations with negative total energy. While there are no prolonged democratic resonances -- phases during which the particles are in approximate energy equipartition \citep{HeggieHut1993} -- in a positive-energy three-body interaction, we contend that the question of binary formation still depends very sensitively on the initial conditions, because the energy exchanges required to form binaries are as large as the total energy, and the general three-body problem is chaotic. Therefore, while one, single scattering problem does not cover all the available phase space, an ensemble of them would, which is the more relevant case for astrophysical applications. The results of this paper can be viewed as a calculation of the distribution of the ergodic subset of three-body binary formation.  

Under these assumptions, we find \citep{GinatPerets2021a} that the outcome probability distribution $f(E_b, \mathbf{S}|E,\mathbf{J})$ for the binary's orbital parameters (those of the outer orbit are automatically given by conservation laws) is
\begin{equation}\label{eqn: outcome distribution}
    f(E_b, \mathbf{S}|E,\mathbf{J}) \propto m_b\frac{\theta_{\max}(R_0,E_b,S)\theta_{\max}(R_s,E_s,\abs{\mathbf{J}-\mathbf{S}}) }{\abs{\mathbf{J}-\mathbf{S}}E_s^{3/2}\abs{E_b}^{3/2}},
\end{equation}
where $E_s = E-E_b$, \emph{etc.}, and $\theta_{\max}$ is a function of its arguments, explicitly defined in \citet[][supplementary information; denoted $l_{\max}$ there]{StoneLeigh2019}. The quantity $R_s$ will be defined in \S \ref{subsec: additional phase-space constraints} below. 
The second factor of $\theta_{\max}$ is the usual one \citep{StoneLeigh2019}, which comes from integrating over the mean anomaly of the outer orbit, explicitly accounting for the finite distance the ejected star can have from the centre of mass of the binary as the triple leaves the strong-interaction region. The first factor of $\theta_{\max}$ has a similar origin: if $E_b>0$, the mean anomaly of the binary is also likewise restricted, by $R_0$: essentially, we require that the separation of the binary be smaller than $R_0$. 
As $\theta_{\max}(R,E,L) \sim R^{-3/2} E^{3/2}$, the first factor of $\theta_{\max}$ cancels the divergence at $E_b =0$, while the second one cancels the divergence of $E_s^{-3/2}$ at $E_s = 0$. 
The parameter $R_s$ is defined as in equation (10) of \cite{GinatPerets2021a} for the $E<0$ case, and in \S \ref{subsec: additional phase-space constraints} below for $E>0$. 

\subsection{Cut-Offs}
\label{subsec:Cut-Offs}
The only thing left to specify is the dependence of $R_0$ on the conserved quantities. We envisage two possibilities for $R_0$: first, one may approximate $R_0$ by the `virial' radius 
\begin{equation}
    R_0 \approx R_{E} \equiv \frac{G(m_1m_2+m_1m_3+m_2m_3)}{6E} = \frac{GM_2^2}{2E},
\end{equation}
where we have defined 
\begin{equation}
    M_2 \equiv \sqrt{\frac{m_1m_2+m_2m_3 + m_1m_3}{3}}.
\end{equation}
This captures the intuition, that the distance of the closest approach of the triple should be such that the potential energy is of the same order of magnitude as the kinetic energy, at least there. Otherwise, a large energy exchange -- necessary for the formation of hard binaries -- is unlikely. Alternatively, the other possibility is to relax this requirement, and simply demand that the distance of the closest triple approach be less than the system's Hill radius, i.e. that at least there, the interaction is a three-body interaction (to leading order), and that the other cluster stars can be neglected. This amounts to 
\begin{equation}
    R_0 \approx d,
\end{equation}
where $d \equiv n^{-1/3}$, with $n$ being the cluster's number density. As mentioned above, $d$ may be defined as the binary's Hill radius in its motion in the cluster, by equating the tidal force that the latter exerts on it with its own gravity, in which case $d \propto M^{1/3}$. For our purposes, what matters is that $d$ is independent of the triple's energy or angular momentum. The possible definitions are
\begin{equation}\label{eqn:R0 definition}
    R_0 = \begin{cases}
        \min\set{R_E,d} & \mbox{ case 1; } \\ 
        d & \mbox{ case 2.} \\ 
    \end{cases}
\end{equation}
While case $2$ might be more justifiable as more generic, the first, more restrictive case is required for significant energy exchanges, if a hard binary is to be produced. 
We will discuss both choices for $R_0$ in this paper.

\subsection{Additional Phase-Space Constraints}
\label{subsec: additional phase-space constraints}
Contrary to the negative-energy three-body problem, it is strictly impossible for a positive-energy triple to exist in a democratic resonance for an extended time. This implies that there could be triple configurations that cannot dynamically arrive at a binary-single system, despite having the right conserved quantities. Hence, one must impose {\emph additional} constraints, to ensure that a triple encounter would be able to form a binary.
The fundamental assumption of this work is that after these extra ones are imposed, the system \emph{is} able to reach all the allowed binary-single configurations. 
Though the above-mentioned various assumptions are not trivial, the successful comparison with the few-body results supports their validity.

While for the negative energy three-body problem, in the case that a hard binary forms, the single star can be ejected at a distance $r_s \leq R_-$ from the centre of mass of the remnant binary, with \citep{GinatPerets2021a}
\begin{equation}
    R_- = \beta \min \set{a_b,\left(\frac{G\mu_b\mu M}{m_b \abs{E}}\right)^{1/3}a_b^{2/3}},
\end{equation}
where we use $\beta = 1.5$ as in \cite{GinatPerets2022}.
However, in our case, there are two complications: (i) the remnant binary might not be hard, i.e. it might have $a_b \gg \frac{G\mu_b\mu M}{m_b \abs{E}}$, and (ii) it could be unbound completely. 

For case (ii), there is no binary formation, which means that there is no restriction on the ejection of star $s$ at all, and in that case, $r_s \leq R_0$ is the only condition. Besides,
for marginally bound binaries (case (ii)), the requirement $r_s \leq R_-$ does not make sense any more, because it stemmed from a requirement of a hierarchy of energies \citep{GinatPerets2021a}, which becomes invalid. Instead, we extend the unbound upper limit to these binaries. For $0 <  \eps \ll 1$, if $\abs{E_b} < \eps Gm_b\mu_b/(2R_0)$, we therefore also only require $r_s \leq R_0$. We choose $\eps = 1/10$, but have verified that the results remain unchanged up to $\eps = 1/100$.  

In summary, the cut-off on the separation $r_s$ between the single star and the centre-of-mass of the other two, when the interaction is deemed to have terminated, is $r_s \leq R_s$, where
\begin{equation}
    R_s = \begin{cases}
        \min \set{R_-, R_0} & \mbox{ if } E_b < -\eps \frac{Gm_b\mu_b}{2R_0} \\ 
        R_0 & \mbox{ otherwise.}
    \end{cases}
\end{equation}
It is this $R_s$ that enters the $\theta_{\max}$ in equation \eqref{eqn: outcome distribution}.


\subsection{Binary Formation For Fixed Conserved Quantities}

\begin{figure}
    \centering
    \includegraphics[width=0.48\textwidth]{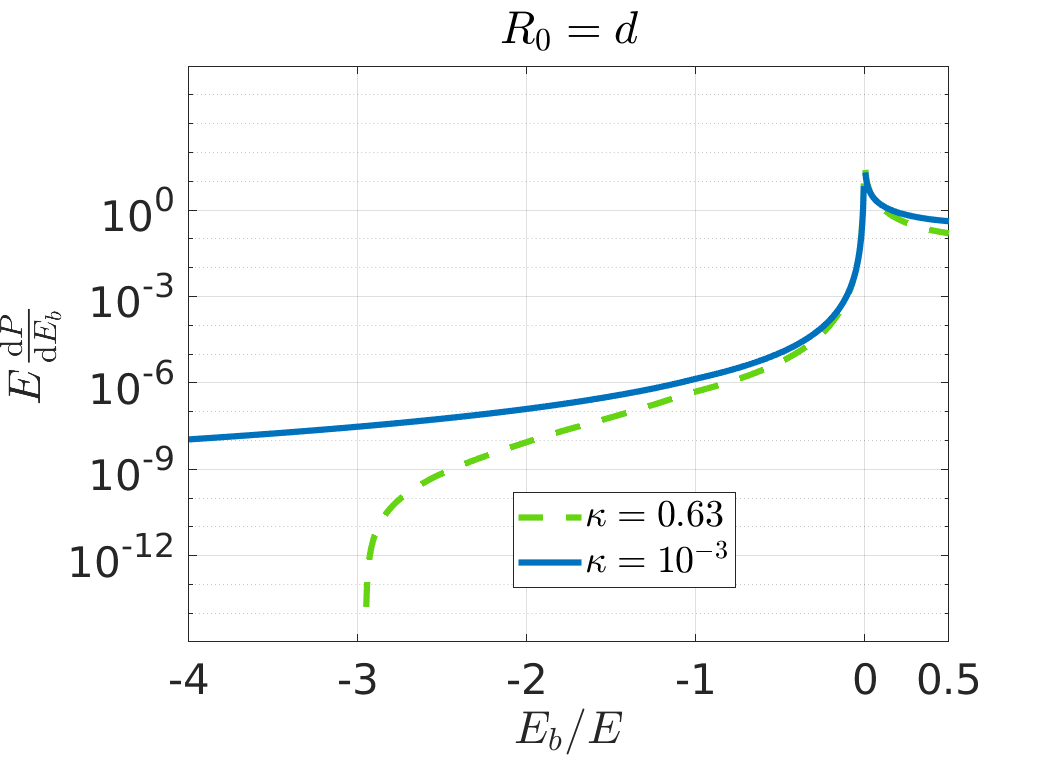}
    \includegraphics[width=0.48\textwidth]{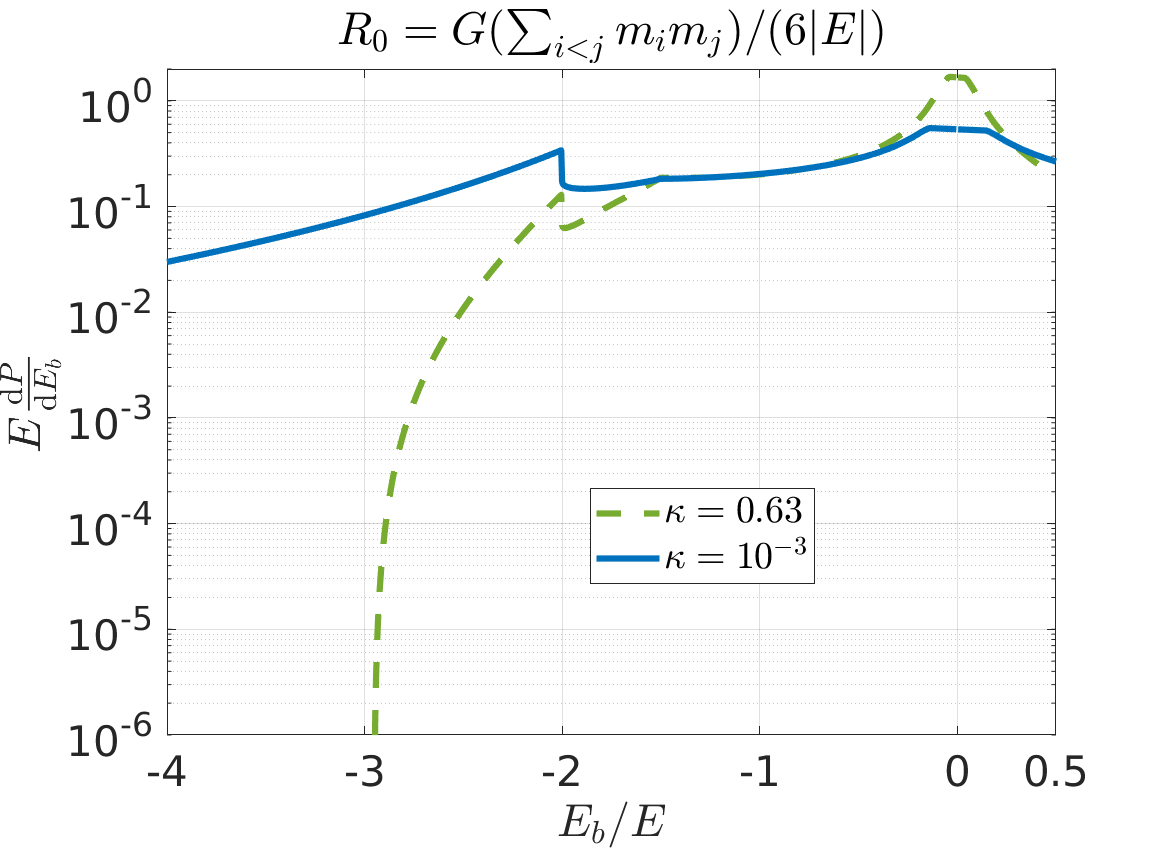}
    \caption{The marginal energy distribution of the remnant binary, for equal masses. Top: $R_0 = d$, with $d = 150 GM\mu/E$. Bottom: the case $R_0 = \min\set{R_E,d}$.}
    \label{fig:energy distribution}
\end{figure}

When evaluated using techniques similar to those of \cite{GinatPerets2021a}, the marginal energy distribution of the remnant binary we find is shown in figure \ref{fig:energy distribution}.

\begin{figure}
    \centering
    \includegraphics[width=0.45\textwidth]{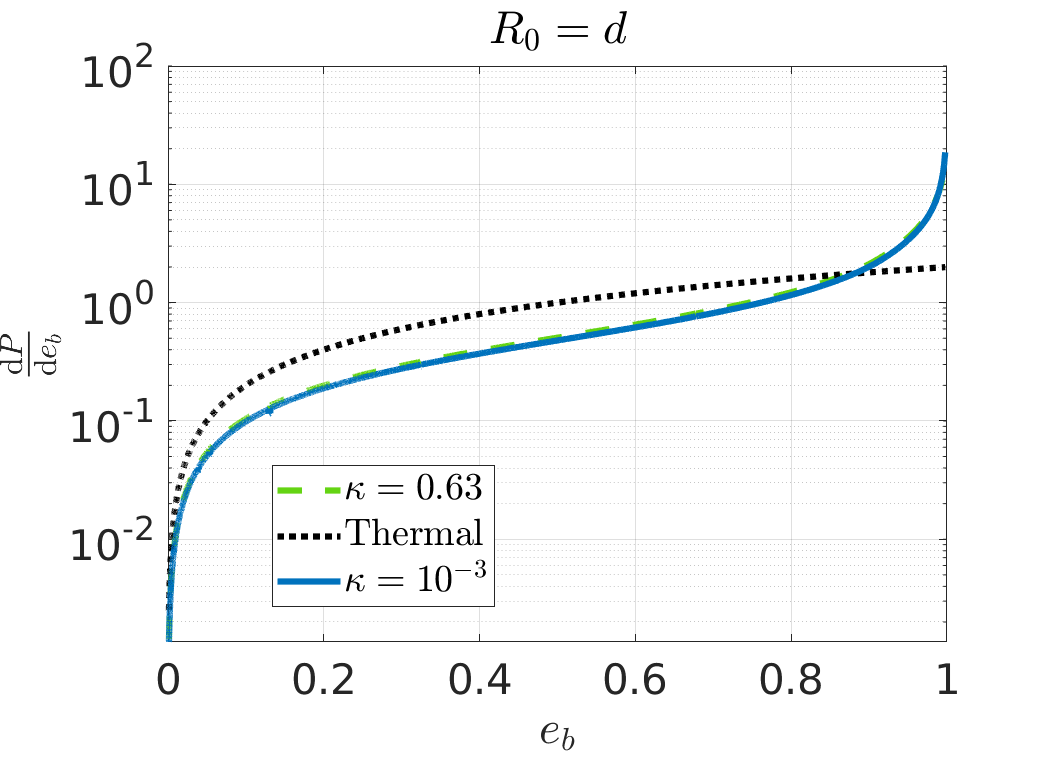}
    \includegraphics[width=0.45\textwidth]{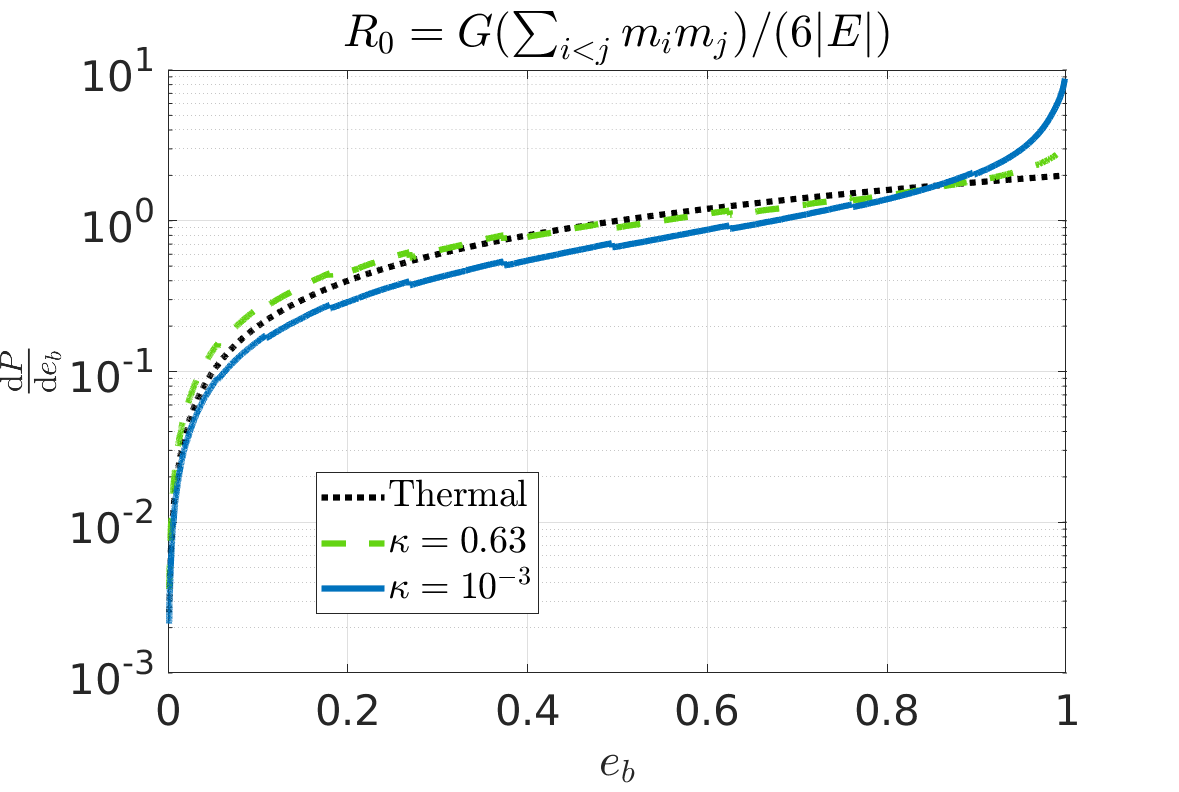}
    \includegraphics[width=0.45\textwidth]{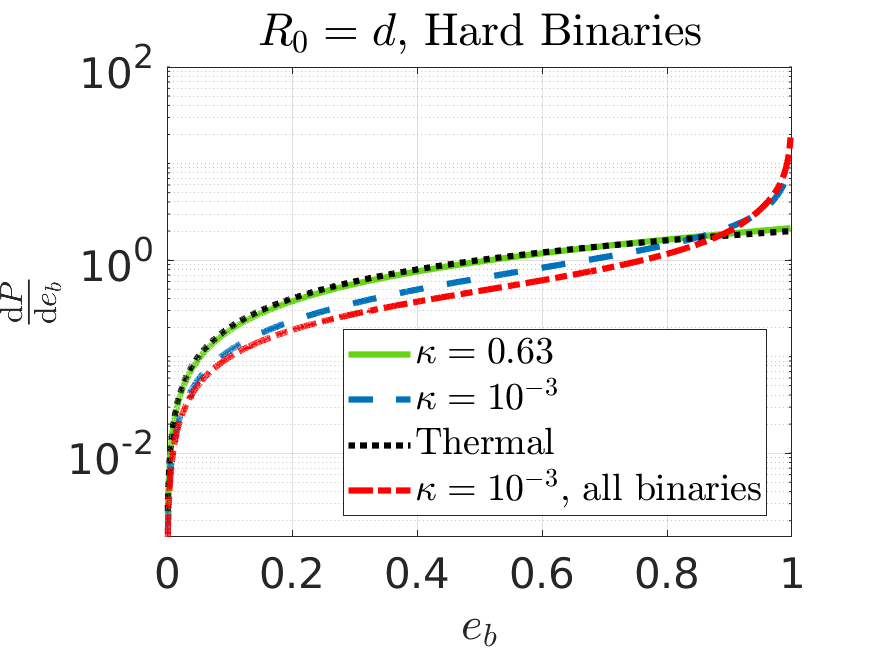}
    \caption{The marginal eccentricity distribution of the remnant binary, given that the remnant binary is bound, for equal masses. This is calculated by integrating equation \eqref{eqn:E_b and S joint distribution} over the domain $E_b<0$ (and normalising). \emph{Top}: $R_0 = d$, with $d = 150 GM\mu/E$; \emph{middle}: the case $R_0 = \min\set{R_E,d}$; \emph{bottom}: same as the top panel, but for hard binaries only, i.e. those with $\abs{E_b} \geq E$.}
    \label{fig:eccentricity distribution}
\end{figure}

\begin{figure*}
	\centering
    \includegraphics[width=0.49\textwidth]{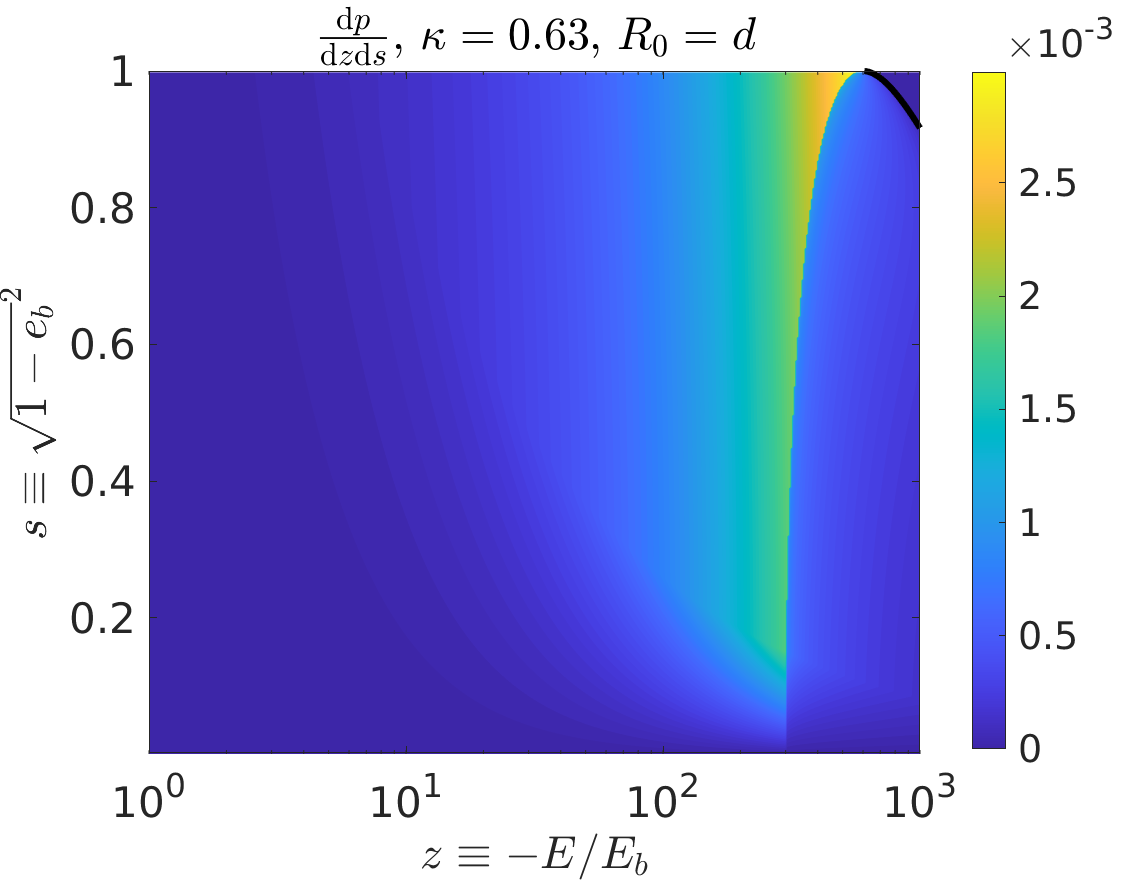}
	\includegraphics[width=0.49\textwidth]{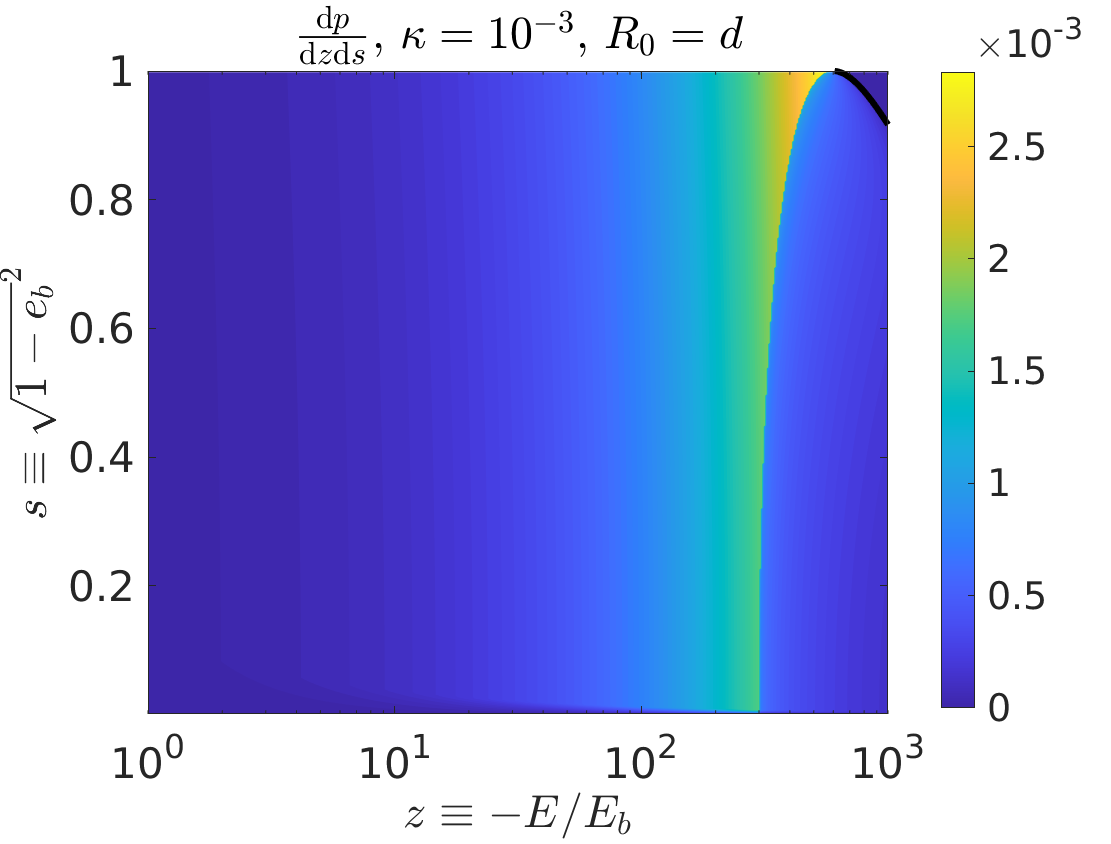}
	\includegraphics[width=0.49\textwidth]{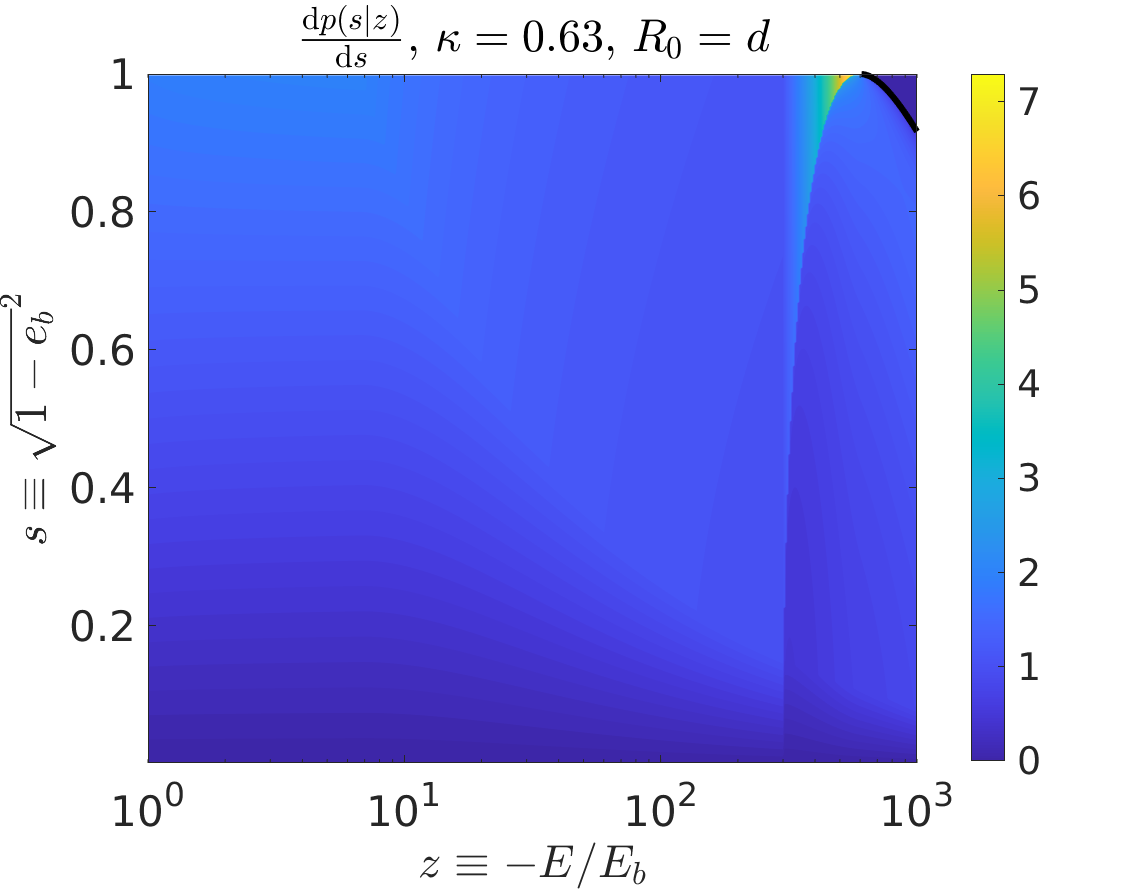}
	\includegraphics[width=0.49\textwidth]{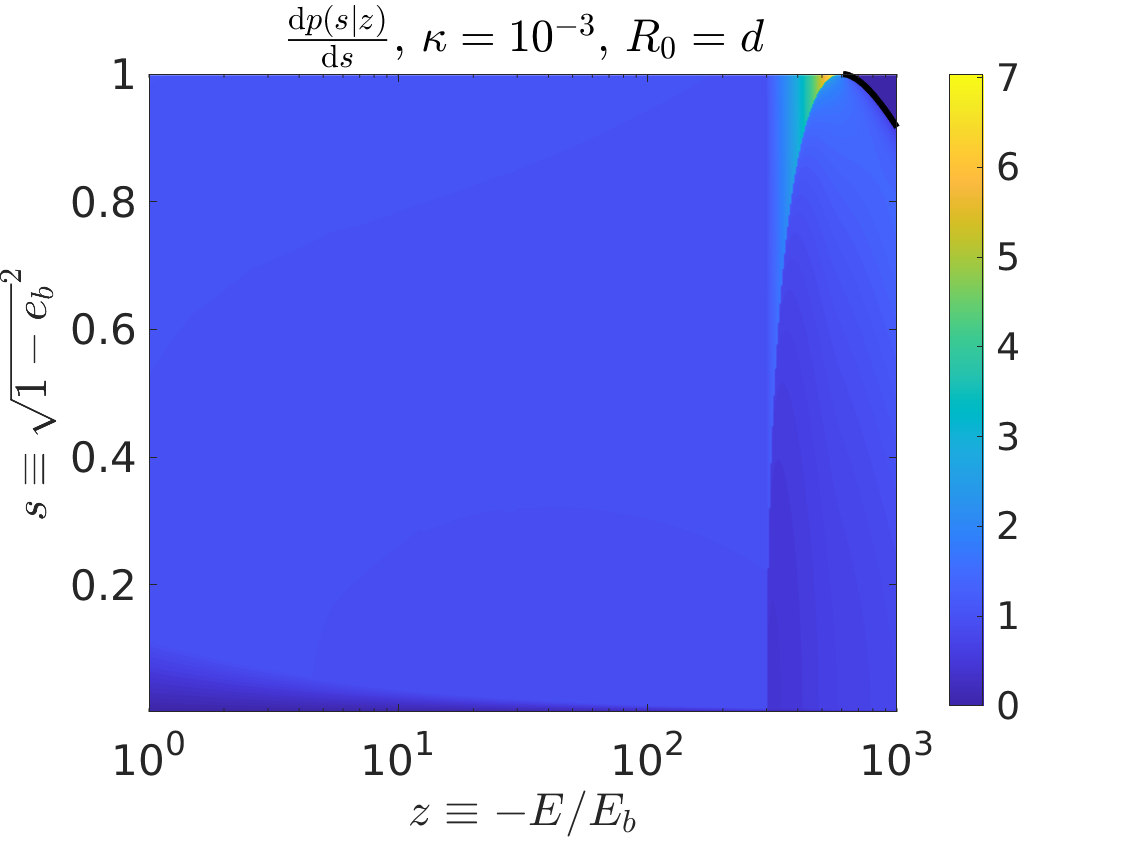}
	\caption{The joint energy-eccentricity distribution of the remnant binary, i.e. equation \eqref{eqn:E_b and S joint distribution}, given that the remnant binary is bound, for equal masses. The plots in the top row show the distribution for different values of the parameter $\kappa = \frac{J^2\abs{E}}{G^2M^2\mu^3}$. The plots in the bottom row are the probability distributions for the dimension-less angular momentum $s \equiv \sqrt{1-e_b^2}$, \emph{given} $z$. The black line is the line $e_b = 1-R_0/a_b$ -- values of $s$ above that threshold cannot form.}
	\label{fig:joint distribution}
\end{figure*}

As only the second $\theta_{\rm max}$ depends on the inclination, one can derive a marginal eccentricity-energy distribution, given that a bound binary forms, in much the same manner as \cite{GinatPerets2022}. The result is 
\begin{equation}\label{eqn:E_b and S joint distribution}
    f_{\rm bd}(E_b, S|E,\mathbf{J}) \propto m_b\frac{\theta_{\max}(R_0,E_b,S)\theta_{ap}(R_s,E_s) }{E_s^{3/2}\abs{E_b}^{3/2}}\frac{A_p \Delta \psi}{J},
\end{equation}
where $\Delta\psi$ and $A_p$ are defined in that work (equation (4) there and the text around it). The marginal eccentricity distribution is plotted in figure \ref{fig:eccentricity distribution}. Due to the extra factor of $\theta_{\max}(R_0,E_b,S)$, the result is highly non-thermal, even for large values of the total angular momentum $J$. Intuitively, the reason that this distribution is super-thermal, is because of the additional factor of $\theta_{\rm max}(R_0,E_b,S)$: this introduces a constraint on the pericentre -- namely that it be smaller than $R_0$, which is of course easier to fulfil when the eccentricity is high; thus, there is more phase-space volume associated with high eccentricity that the binary can occupy. The joint energy-eccentricity distribution is plotted in figure \ref{fig:joint distribution}, for different values of the parameter 
\begin{equation}
    \kappa \equiv \frac{J^2\abs{E}}{G^2M^2\mu^3}.
\end{equation}
One can see immediately that the softest binaries -- those with large $E/\abs{E_b}$ -- tend to be the most eccentric. If the system is in a cluster with temperature $T$, another dimension-less parameter we will require is $\zeta$, defined by 
\begin{equation}\label{eqn:zeta definition}
    \zeta \equiv \frac{GM^2}{R_0k_{\rm B} T}.
\end{equation}

\begin{figure}
    \centering
    \includegraphics[width=0.49\textwidth]{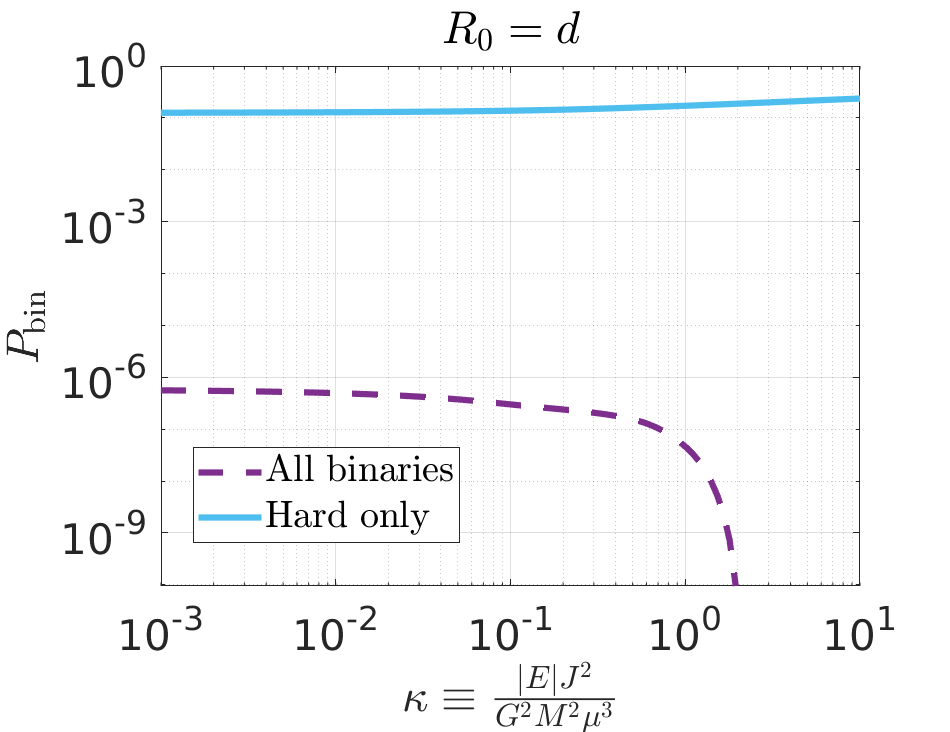}
    \includegraphics[width=0.49\textwidth]{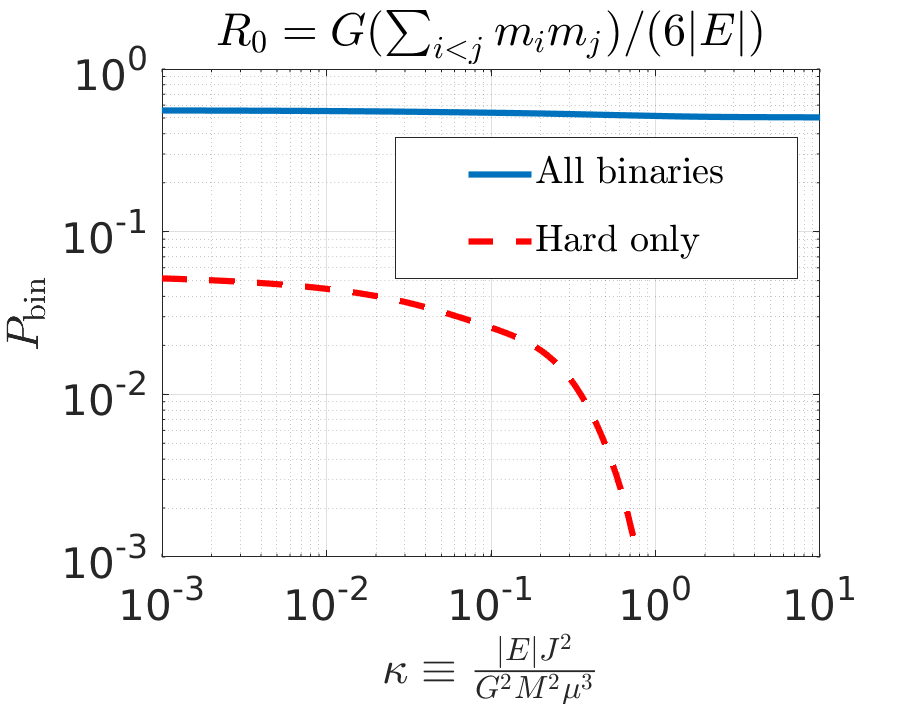}
    \caption{The binary formation probability, as a function of the angular momentum, made dimensionless, for equal masses. \emph{Top}: the case $R_0 = d$, with $d = 150\,GM\mu/E$; \emph{bottom}: the case $R_0 = R_E$. As hard binaries require significant energy exchanges to form, essentially all hard binaries form for $R_0 \leq R_E$, so that their relative fraction is lower when $R_0$ can be larger -- because only soft binaries form at $R_E \leq R_0 \leq d$.}
    \label{fig:binary formation probability}
\end{figure}

The total probability that a binary form is just the integrand of equation \eqref{eqn: outcome distribution} over $E_b < 0$. This is plotted in figure \ref{fig:binary formation probability}, as a function of the only non-dimensional combination of conserved quantities available, for equal masses, i.e. of $\kappa$, for both options for $R_0$ considered here, for a fixed $d$. One can see that when $J$ becomes too large, the probability of hard binary formation plummets to 0. 

An inspection of figure \ref{fig:binary formation probability} shows that $p_{\rm bin}$ changes very little with $\kappa$, from $0.1257$ to $0.2353$ in the range plotted. This is true, of course, only for low values of $\kappa$, because $\kappa$ is bounded from above by Sundman's inequality \citep{Sundman}, which effectively implies that there exists an upper bound $\kappa_{\max}$ above which $p_\bin = 0$. We give explicit expressions for $\kappa_{\max}$ in appendix \ref{appendix: Sundman}. When evaluating the rate in \S \ref{sec:rate} below, we therefore approximate it as a constant, \emph{viz.} 
\begin{equation}\label{eqn:p bin approximate Heaviside}
    p_{\bin}(\kappa) \approx \overline{p}_{\bin}\Theta(\kappa_{\max} - \kappa),
\end{equation}
with $\overline{p}_{\rm bin}$ being the mean value or $p_{\rm bin}(\kappa)$, and $\Theta$ is Heaviside's function. $\overline{p}_{\rm bin}$ still depends on $R_0$, as will be explained below. It is $0.1574$ for $R_0 = 150 GM\mu/E$ in the top panel of figure \ref{fig:binary formation probability}.

\subsection{Hard-Binary Formation}
\label{subsec: hard binary formation}
The probability of the formation of a hard binary is 
\begin{equation}
    p_{\rm hard} = \int_{-\infty}^{-k_{\rm B}T} \mathrm{d}E_b \int dS f_{\rm bd}(E_b,S|E,\mathbf{J}).
\end{equation}
This is closely related to the quantity
\begin{equation}
    p_0 \equiv \int_{-\infty}^{-E} \mathrm{d}E_b \int dS f_{\rm bd}(E_b,S|E,\mathbf{J}).
\end{equation}
For a fixed value of $R_0 = d_0$, $p_0$ depends only on $\kappa$, but $f_{\rm bd}$ scales like $(\textrm{energy})^{-3}$, so its integral scales like $(\textrm{energy})^{-2}$; additionally, $\kappa \propto E$, and therefore 
\begin{equation}
    p_{\rm hard} = \left[\frac{E}{k_{\rm B}T}\right]^2\, p_0\left(\kappa \frac{k_{\rm B}T}{E}\right). 
\end{equation}
We test this scaling in appendix \ref{appendix: p hard and R0}, where we also explain why 
\begin{equation}\label{eqn:scaling p bin with R0}
\begin{aligned}
    & p_{\rm bin} \propto R_0^{-1/2} \\ &
    p_{\rm hard} \propto \frac{G^2M^4}{R_0^2E^2}
\end{aligned}
\end{equation}
in the limit where $\frac{GM^2}{R_0E} \ll 1$. (Let us stress that this is not true for $\frac{GM^2}{R_0E}$ of order unity.) This, and the above, imply that the entire dependence of $p_{\rm hard}$ on the parameters at hand is
\begin{equation}\label{eqn:p hard scaling}
    p_{\rm hard} = \left[\frac{E}{k_{\rm hB}T}\right]^2\left[\frac{d_0}{R_0}\right]^2\, p_0\left(\kappa \frac{k_{\rm B}T}{E};R_0 = d_0\right). 
\end{equation}
This scaling behaviour allows us to calculate $p_{\rm hard}$ for one set of values of the parameters $R_0$ and $k_{\rm B}T$, tabulate it and use that when computing the hard binary formation probability at any value of the parameters, and any $\kappa$. This will simplify the rate calculation in the next section considerably.

\section{Binary Formation Rate Calculation}
\label{sec:rate}
So far, we have explored the binary formation probability \emph{for fixed} total energy and angular momentum. In a realistic cluster, of course, these vary from one interaction to another, and a more relevant quantity is the average value of $p_{\rm bin}$, over an ensemble of triples, with a certain distribution of initial conditions. The purpose of this section is to find this probability, which we call $P_{\rm bin}$, and to characterise it. 


The rate calculation follows \cite{AarsethHeggie1976} (see also \citealt{GoodmanHut1993,Ivanovaetal2005}). The rate per star, at which a star of $m_1$ becomes (on average) involved in a 3-body interaction leading to the formation of a binary with component masses $m_1$ and $m_2$, is given by
\begin{equation}\label{eqn:rate per star}
\begin{aligned}
    \Gamma & = \frac{n_2n_3(m_1m_2m_3)^{3/2}}{(2\pi k_{\rm B} T)^3M^{3/2}}\int\mathrm{d}^3v \int\mathrm{d}^3V\int\mathrm{d}r\int\mathrm{d}R \\ &
    2\pi r v~ 4\pi R^2\mathrm{e}^{-\frac{\mu_bv^2}{2k_{\rm B}T}-\frac{\mu V^2}{2k_{\rm B}T}}p_{\bin}.
\end{aligned} 
\end{equation}
The integration domain for $R$ and $r$ depends on the initial condition distribution. Here, all stars are assumed to have a Gaussian velocity distribution with temperature $T$, and the triple's centre-of-mass motion has already been integrated out. The Jacobi coordinates are 
\begin{align}
    \mathbf{R} & \equiv \mathbf{x}_3-\frac{m_1\mathbf{x}_1 + m_2\mathbf{x}_2}{m_b} \\ 
    \mathbf{r} & \equiv \mathbf{x}_2 - \mathbf{x}_1,
\end{align}
and $\mathbf{V}$ and $\mathbf{v}$ are their respective associated velocities, where $\mathbf{x}_n$ is the position of star $n$. 

The rate per unit volume (instead of per star) is simply equation \eqref{eqn:rate per star} multiplied by $n_1$, i.e. 
\begin{equation}\label{eqn:rate per volume}
\begin{aligned}
    \Gamma_v & = \frac{n_1n_2n_3(m_1m_2m_3)^{3/2}}{(2\pi k_{\rm B} T)^3M^{3/2}}\int\mathrm{d}^3v \int\mathrm{d}^3V\int\mathrm{d}r\int\mathrm{d}R \\ &
    2\pi r v~ 4\pi R^2\mathrm{e}^{-\frac{\mu_bv^2}{2k_{\rm B}T}-\frac{\mu V^2}{2k_{\rm B}T}}p_{\bin}.
\end{aligned} 
\end{equation}

By the law of total probability, $\Gamma_v$ factories into a product of the total interaction rate, $\Gamma_{\rm 3UB}$, and the probability that a binary forms, given a three-body interaction, $P_{\rm bin}$. The former evaluates to \citep{Atallahetal2024} 
\begin{equation}
    \Gamma_{\rm 3UB} = \frac{8\pi^{3/2}}{3\sqrt{2}} R_0^5 n_1 n_2 n_3  \sigma,
\end{equation}
where it is assumed that the bodies have random positions, uniformly distributed in a sphere of radius $R_0$, and a Gaussian velocity distribution, with a $1$-dimensional dispersion $\sigma$. 

Then, $\Gamma_v = \Gamma_{\rm 3UB}P_{\rm bin}$, where
\begin{equation}\label{eqn: probability full}
	\begin{aligned}
		P_{\rm bin} & = \frac{1}{V_0^2}\left(\frac{m_1m_2m_3}{M}\right)^{3/2}\frac{1}{(2\pi k_{\rm B}T)^{3}}\int \mathrm{d}^3r\mathrm{d}^3R\mathrm{d}^3v\mathrm{d}^3V \\ & 
		\mathrm{e}^{-\frac{E}{k_{\rm B}T}} p_{\rm bin}(\kappa(E,J);R_0).
	\end{aligned}
\end{equation}

As mentioned above, the binary formation probability, $p_{\rm bin}$, \emph{given} a triple interaction, is only a function of $\kappa$ and $R_0$ in the equal-mass case, but may also depend on the mass-ratios if they are unequal. At any rate, we calculate it directly by integrating equation \eqref{eqn: outcome distribution} over the allowed parameter space (for a bound remnant binary), summing over the three possibilities for star $s$, and normalising. 

A perhaps more interesting quantity is obtained by replacing $p_\bin$ by the probability for creating a \emph{hard} binary, $p_{\rm hard}$, which gives the rate of formation of hard 3-body binaries, which we denote by $P_{\rm hard}$. The non-trivial shape of $p_{\rm hard}(\kappa)$ (see figure \ref{fig:binary formation probability}) requires a more complicated calculation; both calculations, for $P_{\rm hard}$ and $P_{\rm bin}$ are performed in appendix \ref{appendix: rate calculation}. 
We restrict the calculations to the $\zeta \ll 1$ r\'{e}gime, which is the physically relevant one. We now move on to describe the integration domain and derive $P_{\rm bin}$, $P_{\rm hard}$, the rate, and the orbital parameter distribution.

\subsection{Occurrence of an Interaction}
As alluded to above in \S \ref{sec: f_bin for fixed kappa}, for a binary to form, one of the stars must pass sufficiently close to the centre of mass of the other two, to exchange enough energy with them to make them bound. This is not encapsulated in the $\kappa$ dependence of $p_{\rm bin}$ $\emph{per se}$ and must be imposed explicitly when evaluating the integrals for $P_{\rm bin}$ and $P_{\rm hard}$ because while the outcome does depend sensitively on the initial conditions, it does not only depend on them via $\kappa$. 

We can derive a necessary condition for this to happen, based on the minimum requirement of a strong enough perturbation in the impulsive limit (cf. \citealt{AarsethHeggie1976}). In the time-reversed configuration, where a single star ionises a binary, it is required, at a minimum, that the star would be able to give the binary enough energy to unbind it. That is, in the impulse approximation (which is a lower limit)
\begin{equation}
    \sqrt{\frac{Gm_b}{a_b}}\frac{2Gm_s}{r_sV} \geq \frac{Gm_b}{2a_b};
\end{equation}
this should be multiplied by $\sqrt{2}$ because both stars are affected. Hence,
\begin{equation}\label{eqn:impulse}
    r_s \leq  4\sqrt{\frac{Gm_s^2\mu_s a_b}{E_s m_b}}.
\end{equation}
We therefore restrict $R$ as follows: firstly, $R \leq R_0$, and additionally, 
\begin{equation}\label{eqn:cut-offs initial}
    R \leq \begin{cases}
        \sqrt{8\frac{G^2m_s^2\mu_s\mu_b}{E_s^iE_b^i}} & \mbox{ hard only} \\ 
        \frac{2Gm_s}{V}\sqrt{\frac{r}{2Gm_b}} & \mbox{ all binaries},
    \end{cases}
\end{equation}
where here $E_s^i = \mu_sV^2/2$ and $E_b^i = \mu_b v^2/2$. The first case is the same as \eqref{eqn:impulse}; the second case is less stringent because, for the formation of a soft binary, we require merely that one of the stars impart an impulse on the other two, which is as large as their relative momentum at (hyperbolic) pericentre. Whether subsequent interactions in the same encounter then form a binary is determined by $p_{\rm bin}$. 

\subsection{Computation of The Formation Probability}
We are now in a position to evaluate the integrals in equation \eqref{eqn: probability full}, both for $P_{\rm bin}$ and for the analogous $P_{\rm hard}$. This is done in appendix \ref{appendix: rate calculation}, where we also show analytically that, as functions of $\zeta$, 
\begin{equation}\label{eqn:scaling laws for the full probability}
\begin{aligned}
	P_{\rm bin} & \sim P_{\rm bin}(\zeta_0) \left(\frac{\zeta}{\zeta_0}\right)^2 \propto \chi_1^{-2}, \\ 
	P_{\rm hard} & \sim P_{\rm hard}(\zeta_0) \left(\frac{\zeta}{\zeta_0}\right)^5\propto \chi_1^{-5},
\end{aligned}
\end{equation}
as $\zeta \to 0$.
In particular, this implies that as $R_0 \to \infty$, both $\Gamma$ and $\Gamma_v$ tend to a constant, for hard binaries, but grow as $R_0^3$ if one wishes to include soft binaries, too. 

\begin{figure}
	\centering
	\includegraphics[width=0.49\textwidth]{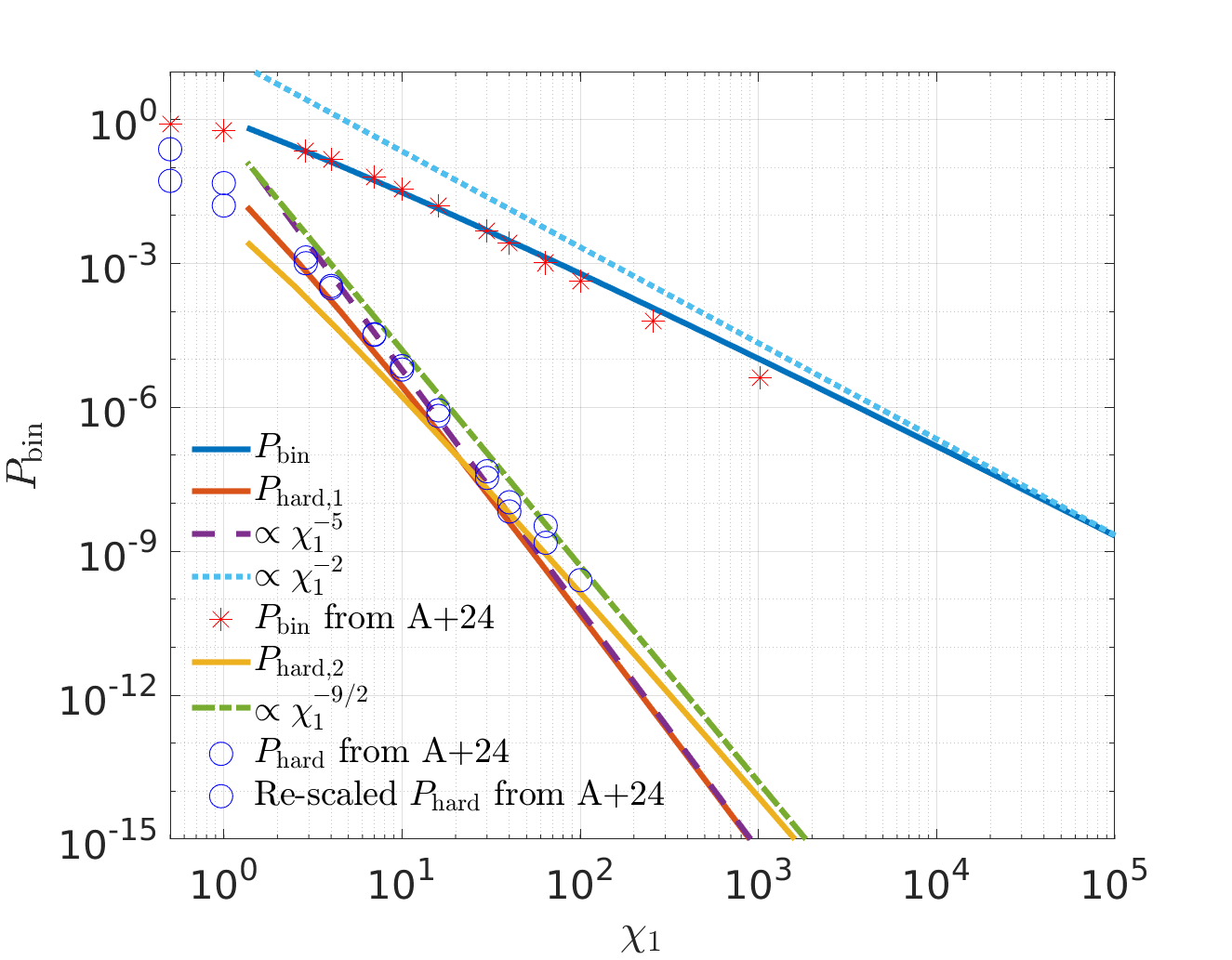}
	\caption{The binary formation probability, as a function of $\chi_1 = 27/(2\zeta)$, for equal masses. This assumes a density $n = 10^5 ~\textrm{pc}^{-3}$, $m = 1~M_\odot$, and a velocity dispersion $\sigma = 10 \textrm{km s}^{-1}$. Data from the numerical experiments of \citet[figure 3]{Atallahetal2024} are over-plotted for comparison., and re-scaled according to their equation (20). Details are in the text and appendix \ref{appendix: rate calculation}.}
	\label{fig:probability}
\end{figure}

We show the full results of calculating $P_{\rm bin}$ and $P_{\rm hard}$ as functions of $\zeta$, for the equal mass case in figure \ref{fig:probability}. We plot them as functions of a dimension-less parameter $\chi_1 \propto 1/\zeta$, which is defined in \cite{Atallahetal2024}, for ease of comparison. In the equal mass case, $\chi_1 = 27/(2\zeta)$. We see good agreement between our results and those of \cite{Atallahetal2024}.\footnote{The procedure whereby the initial conditions are changed in that work, as the particles are moved backward along their initial velocity vectors, changes neither the angular momenta nor, in the limit $\zeta \ll 1$, the energies, so we disregard it here.} Surprisingly, the asymptotic $\chi_1$ dependence is not reached until very large values of $\chi_1$. The two curves, $P_{\rm hard,1}$ and $P_{\rm hard,2}$ are plotted when using the first of conditions \eqref{eqn:cut-offs initial} or the second for hard binaries, respectively (equations \eqref{eqn: total p hard} and \eqref{eqn: total p hard 2}). 

It is also instructive to compare the total hard binary formation rate to $\Gamma_v$, to the results of previous works: \cite{GoodmanHut1993} (cf. \citealt{HeggieHut2003}) obtain, for the equal mass case,
\begin{equation}
	\Gamma_{\rm GH} = 0.75 \frac{G^5m^5 n^3}{\sigma^9}.
\end{equation}
We find 
\begin{equation}
	\lim_{\zeta \to 0} \Gamma_v = 0.68 \Gamma_{\rm GH};
\end{equation}
that is, the contribution of ergodic hard binary formation to the total rate is roughly $70\%$, or
\begin{equation}\label{eqn:total hard binary formation rate 1}
	\Gamma_v = 0.51 \frac{G^5m^5 n^3}{\sigma^9}.
\end{equation}

\subsection{Orbital Parameter Probabilities}
\begin{figure*}
	\centering
	\includegraphics[width=0.49\textwidth]{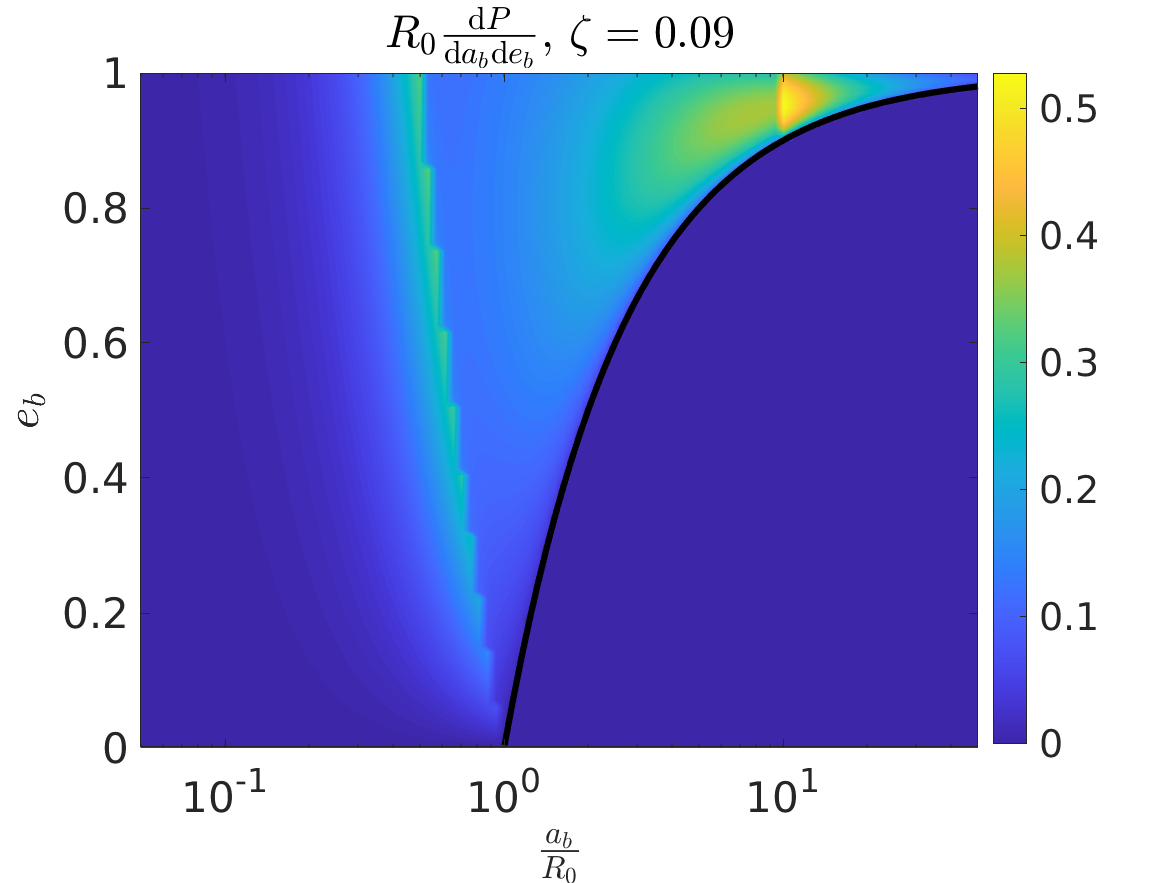}
    \includegraphics[width=0.49\textwidth]{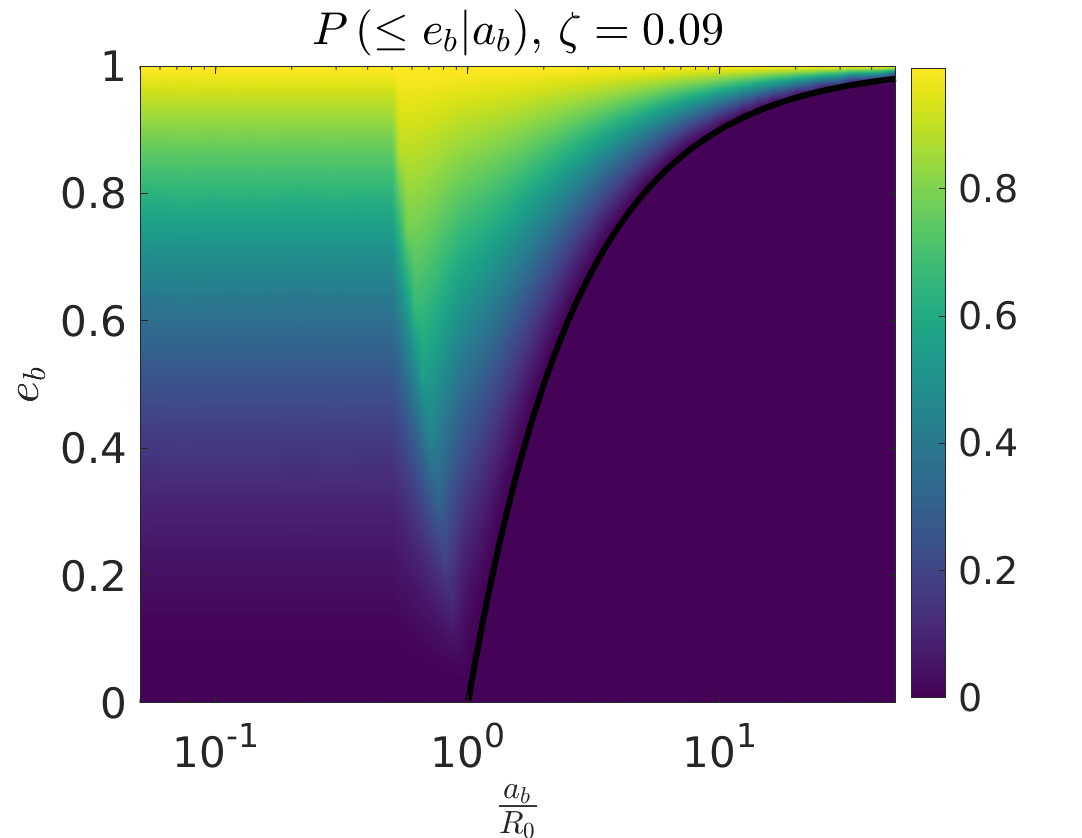}
    \includegraphics[width=0.49\textwidth]{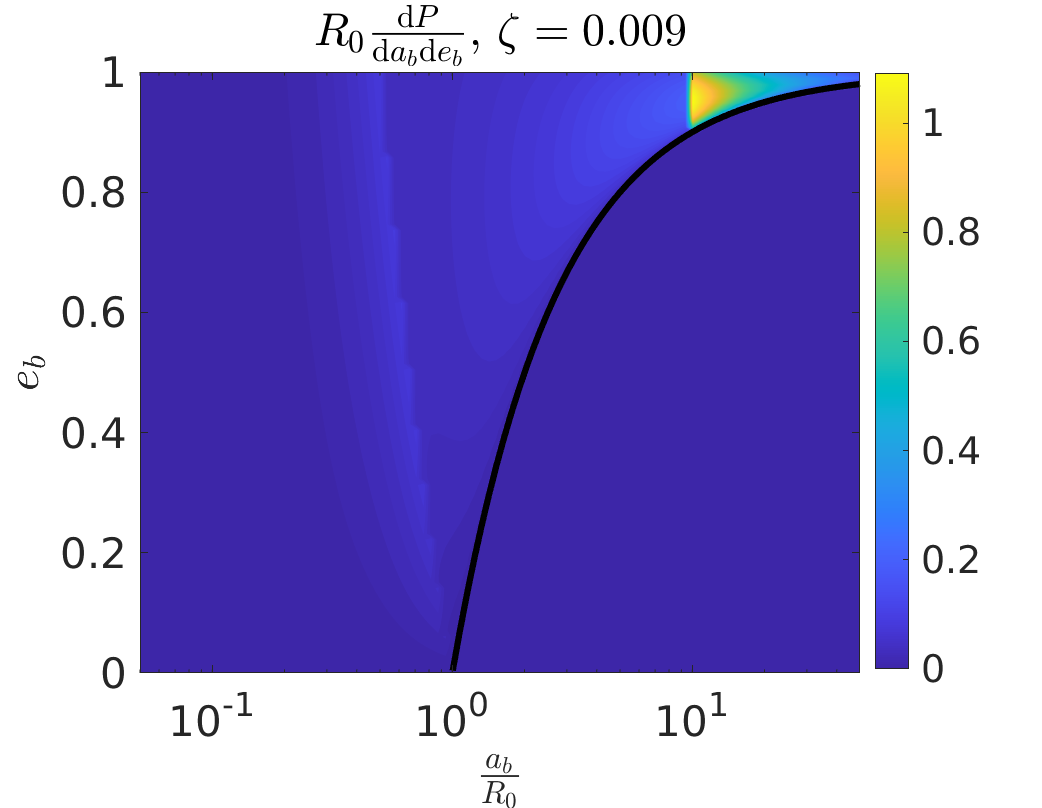}
    \includegraphics[width=0.49\textwidth]{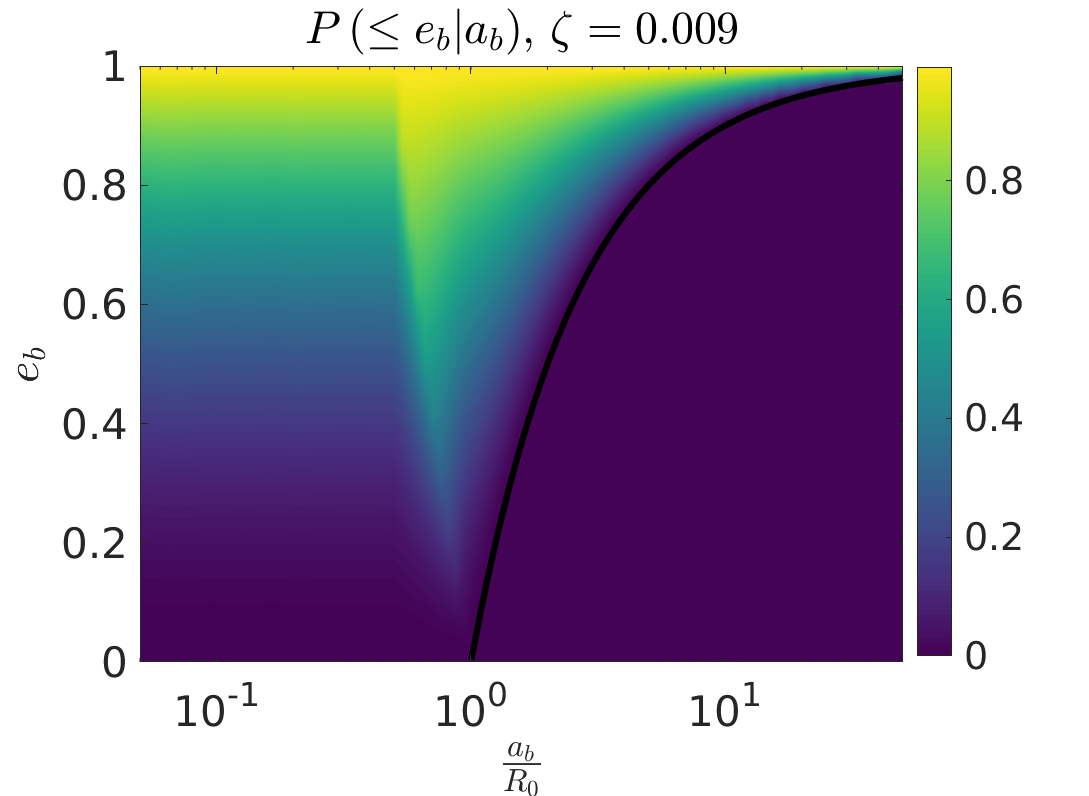}
	\caption{Left: The joint probability distribution of the semi-major axis and the eccentricity of three-body binaries in a cluster. Right: the cumulative eccentricity distribution, given a semi-major axis. These are evaluated for equal masses at $\zeta = 0.09$ (top) and $\zeta = 9\times 10^{-3}$ (bottom). The black line is $e_b = 1-R_0/a_b$. While the joint distributions depend on $\zeta$, the conditional eccentricity distribution does not seem to do so, in the $\zeta \ll 1$ limit. See text for more details.}
	\label{fig:probability of a and e}
\end{figure*}

One can play the same exercise, and compute the total formation probability of binaries with given values of the semi-major axis and eccentricity, $(a_b,e_b)$, \emph{given} that a binary does form. This is a convex combination of functions of the form \eqref{eqn:E_b and S joint distribution}, with $E$ and $J$ sampled from the distribution of initial conditions described above. That is, 
\begin{equation}\label{eqn: P of a and e in a cluster}
	\begin{aligned}
		\frac{\mathrm{d}P}{\mathrm{d}a_b\mathrm{d}e} & = \frac{1}{P_{\rm bin}}\left[\frac{m_1m_2m_3}{M}\right]^{3/2}\frac{V_0^{-2}}{[2\pi k_{\rm B}T]^{3}}\int \mathrm{d}^3r\mathrm{d}^3R\mathrm{d}^3v\mathrm{d}^3V \\ & 
		\mathrm{e}^{-\frac{E}{k_{\rm B}T}} ~f_{\rm bd}(E_b,S|E,J;R_0)\frac{\mathrm{d}E_b}{\mathrm{d}a_b}\frac{\partial S}{\partial e_b},
	\end{aligned}
\end{equation}
where we have made the $R_0$ dependence of $f_{\rm bd}$ explicit. We evaluate this $6$-dimensional integral using Monte Carlo integration. The result is shown in figure \ref{fig:probability of a and e}. Also shown in that figure is the line $e_b = 1-R_0/a_b$, below which there is zero probability of forming binaries. This effectively forces the soft binaries, with $a_b \geq R_0$, to be very eccentric. To compare with figure 7 of \cite{Atallahetal2024} we also plot the cumulative distribution $P(\leq e_b|a_b)$. This is calculated by integrating the probability density in the left panel of figure \ref{fig:probability of a and e} over eccentricity up to $e_b$, and normalising, at each value of $a_b$. The result is shown in the right panel of figure \ref{fig:probability of a and e}. One can see that the two colour maps agree qualitatively, in both $a_b \ll R_0$ and $a_b \gg R_0$ limits, although our $P(\leq e_b|a_b)$ has a ridge which is not present in the simulation results of \cite{Atallahetal2024}; at present, we do not know what gives rise to this ridge. 

\section{Application}
\label{sec: application}
To summarise, let us describe briefly how to apply the results of this paper to modelling of binary populations in a cluster with a given temperature and density: first, one should calculate $\zeta$ with equation \eqref{eqn:zeta definition}. Then, compute $P_{\rm bin}$ with equation \eqref{eqn: total p bin} and $P_{\rm hard,1,2}$ with equations \eqref{eqn: total p hard} and \eqref{eqn: total p hard 2}. The rate is then $\Gamma_{3\textrm{UB}}$ times the probability. For $P_{\rm hard,1}$, if $\zeta \ll 1$ the hard binary formation rate is directly given by 
\eqref{eqn:total hard binary formation rate 1}. 

However, one should bear in mind that the orbital parameters $a_b$ and $e_b$ of the binaries that form are highly correlated, and their distribution is also highly dependent on $\zeta$. The conditional probability, $P(e_b|a_b)$ does not appear to depend strongly on $\zeta$ from figure \ref{fig:probability of a and e}, and is thus independent of the environment in the astrophysically realistic limit $\zeta \ll 1$. In creating a realistic distribution of three-body binaries one must, therefore, draw them from the joint distribution $P(a_b,e_b)$ in figure \ref{fig:probability of a and e}. For these one has to use equation \eqref{eqn:E_b and S joint distribution} and the procedure at the end of \S \ref{sec:rate}, or equivalently read these off figure \ref{fig:probability of a and e}. 

\section{Discussion and Summary}
\label{sec:discussion}

This paper consisted of two main parts: the first is concerned with the three-body problem, giving a probability distribution for the outcome of a three-body scattering experiment, for fixed values of the conserved quantities, i.e. $E$, $\mathbf{J}$ and the masses.
The second part of the paper was an application of the first to a cluster -- a system with a certain distribution of conserved quantities (in the equal mass case). This distribution was a Gaussian velocity distribution, while initial positions were assumed to be uniform in a sphere of radius $R_0$. 

Everything done here assumed some kind of ergodicity or phase-space mixing. The main argument made to support this was that while indeed there are no democratic resonances for an extended period of time, the problem is still chaotic, and whether a binary forms and its final characteristics, do depend very sensitively on the initial conditions because the binding energies required for a binary are much smaller than $E$. Alternatively, at least for the negative-energy problem, we know that detailed balance arguments produce the same probability distribution as the one obtained from ergodic mixing in the available phase-space volume \citep{GinatPerets2021a}. Here, though, we do not find the same result as \cite{GoodmanHut1993}, which was based on detailed balance, primarily because that work did not account explicitly for angular momentum conservation, but also because in the negative-energy case, the balance was between bound triples (in a democratic resonance) and binaries and singles, while here there are no democratic resonances. 

However, this can also be viewed from the opposite perspective, where the rates that we calculated here are those for three-body binary formation via an ergodic channel. They were slightly lower than those computed from numerical simulations (cf. figure \ref{fig:probability}), possibly due to non-ergodic three-body binary formation also playing some r\^{o}le, or due to inexact matching of the initial condition distribution. But the agreement, albeit imperfect -- and the correct scalings with $\zeta$ -- act as an \emph{a posteriori} validation of the ergodicity assumption. 

We found that soft binaries were formed at a much higher rate and that their eccentricity distribution tended to be highly non-thermal (this arose primarily from the restriction $e_b \geq 1-R_0/a_b$), while hard binaries had a thermal distribution, except at low angular momenta. When $\kappa \approx 0$, the zero angular momentum case for the three-dimensional three-body problem was already known to be very similar to the two-dimensional problem \citep{ValtonenKarttunen2006,StoneLeigh2019,Parischewskyetal2021,Samsingetal2022,GinatPerets2022,Tranietal2024,FabjSamsing2024}, which also yields super-thermal eccentricity distributions. That soft binaries are highly preferred (which is corroborated by \citealt{GoodmanHut1993,Atallahetal2024}) suggests that these binaries are important in clusters, despite being mostly ignored in the literature,\footnote{See, e.g. \cite{RoznerPerets2023} for a statistical study of soft binaries in equilibrium.} which is why the findings of this paper are not in disagreement with those of, e.g., \cite{Gelleretal2019}, who observed a sub-thermal distribution of binaries in a cluster Monte Carlo simulation but with only hard binaries tracked.
Indeed, the minuscule minority of soft binaries that do survive encounters with single stars, and become hard, might not be negligible in comparison with the number of the three-body binaries that form hard at the outset. Moreover, if any gas is present, it is suggested in \cite{RoznerPerets2024} that dynamical friction from it could counteract the softening effect, and shield some soft binaries from destruction.  

That three-body binary formation is a natural mechanism for generating eccentric wide binaries might bear on the recent conundrum of eccentric wide binaries detected in \emph{Gaia} \citep{Tokovinin2020,Hamilton2022,Hwangetal2022a,Hwangetal2022b,HamiltonModak2023,ModakHamilton2023}, although those binaries are not globular cluster ones.

One of the key questions raised by \cite{Atallahetal2024} is the scaling of $P_{\rm hard}$ with $\zeta$ (equivalently $\chi_1$). In a typical scattering experiment (e.g. if gravity was a short-distance force), one would expect that the rate of binary formation converges as the maximum impact parameter tends to infinity. This would require $P_{\rm hard} \sim \zeta^{5}$, which is what we find if we use the first of equations \eqref{eqn:cut-offs initial} to calculate the binary formation rate ($P_{\rm hard,1}$ in figure \ref{fig:probability}). However, \cite{Atallahetal2024} fit a shallower, $\zeta^{9/2}$ scaling to their results, which incidentally is what we find if we also use the second condition of \eqref{eqn:cut-offs initial} also for hard binaries ($P_{\rm hard,2}$ in figure \ref{fig:probability}); this would imply that the total binary formation rate would tend to infinity as $R_0 \to \infty$. One should observe, though, that it is unclear which of $P_{\rm hard,1}$ and $P_{\rm hard,2}$ follows the scaling of the simulation results better, and that in both cases, the asymptotic scaling is achieved only at very large values of $\chi_1$, for which there aren't any numerical data. Additionally, $\zeta$ (equivalently $R_0$) is a physical parameter, which is not infinite in real astrophysical systems -- it is very small, but finite, determined by the inter-particle separation (or by the Hill radius at the triple's position in the cluster), and the local velocity dispersion there.


\section*{Acknowledgements}
We thank Dany Atallah, Alessandro Trani, and Newlin Weatherford for helpful discussions and for sharing their data, used for figure \ref{fig:probability}. This work was supported by a Leverhulme Trust International Professorship Grant (No. LIP-2020-014). Y.B.G. was supported in part by the Simons Foundation via a Simons Investigator Award to A.A. Schekochihin. The colour-map for the right panel of figure \ref{fig:probability of a and e} uses the \verb|Matlab| function \verb|viridis.m| by S. Cobeldick, available \href{https://www.mathworks.com/matlabcentral/fileexchange/62729-matplotlib-perceptually-uniform-colormaps?s_tid=srchtitle}{here}.

\section*{Data Availability}
No new data were produced for this article. Scripts used to make the figures in this article will be shared upon reasonable request to the corresponding author.




\bibliographystyle{mnras}
\bibliography{encounters,my_papers}



\appendix
\section{Limits From Sundman's Inequality}
\label{appendix: Sundman}
Sundman's inequality states that $J^2 \leq 2IK$, with $I$ equal to the triples moment of inertia, $I \equiv \sum_i m_i r_i^2$, and $K$ denoting the total kinetic energy. We approximate $I \approx MR_0^2/3$. 
If $E$ is of the same order as the cluster's thermal energy, $k_{\rm B}T$, then for $\zeta \ll 1$, $K \approx E$, and then Sundman's inequality becomes $\kappa \leq \frac{2}{3}(\zeta k_{\rm B}T/E)^{-2}(M/\mu)^3$. For $\zeta \gg 1$, when the triple has mutual separations $R_0$, $K \approx 3GM_2^2/R_0$, whence Sundman's inequality becomes $\kappa \leq 2M_2^2M/\mu^3 (\zeta k_{\rm B}T/E)^{-1}$. We join the two cases continuously, i.e. by
\begin{equation}
    \kappa \leq \kappa_1 \equiv \frac{2M^3}{\mu^3}\begin{cases}
        \frac{1}{3}\left(\frac{E}{\zeta k_{\rm B}T}\right)^{2} & \mbox{ if } \frac{\zeta k_{\rm B}T}{E} \leq \frac{M^2}{3M_2^2}, \\
        \frac{M_2^2}{M^2}\frac{E}{\zeta k_{\rm B}T} & \mbox{ otherwise.}
    \end{cases}
\end{equation}
In the case of unequal masses, another bound can be useful: 
\begin{equation}
    \kappa \leq \frac{E}{G^2M^2\mu^3}\left(L_{\max}^2+S_{\max}^2+2L_{\max}S_{\max}\right).
\end{equation}
Approximate upper limits for the inner and outer angular momenta are $L_{\max} \lesssim \mu V R_0$ and $S_{\max} \lesssim \mu_b v R_0$, whence 
\begin{equation}
    \kappa \lesssim \kappa_2 \equiv \frac{2M^2}{\mu^2}\frac{E}{\zeta k_{\rm B}T} \left(\frac{E_s \mu + E_b \mu_b + 2\sqrt{\mu\mu_bE_sE_b}}{\zeta \mu k_{\rm B}T}\right).
\end{equation}
We define $\kappa_{\max} \equiv \min\set{\kappa_1,\kappa_2}$. 

For equal masses this simplifies to 
\begin{equation}\label{eqn: kappa max equal masses}
\begin{aligned}
    & \kappa_1 = \frac{243}{4}\begin{cases}
        \left(\frac{E}{\zeta k_{\rm B}T}\right)^{2} & \mbox{ if } \frac{\zeta k_{\rm B}T}{E} \leq 3, \\
        \frac{1}{3}\frac{E}{\zeta k_{\rm B}T} & \mbox{ otherwise.}
    \end{cases} \\ 
    & \kappa_2 = \frac{81}{2}\frac{E}{\zeta k_{\rm B}T}\left[\sqrt{\frac{E_s}{\zeta k_{\rm B}T}} + \sqrt{\frac{3E_b}{4\zeta k_{\rm B}T}} \right]^2 \\ 
    & \kappa_{\max} = \min\set{\kappa_1,\kappa_2}.
\end{aligned}
\end{equation}

\section{Scaling of The Hard Binary Formation Probability With Parameters}
\label{appendix: p hard and R0}
The two parameters $p_{\rm hard}$ can depend on -- in addition to $\kappa$ -- are the ratio $E/(k_{\rm B}T)$ and the ratio $R_0E/(GM^2)$ (in the equal mass case). We explained already in \S\ref{subsec: hard binary formation} why 
\begin{equation}
    p_{\rm hard} \propto \left[\frac{E}{k_{\rm B}T}\right]^2\, p_0\left(\kappa \frac{k_{\rm B}T}{E}\right),
\end{equation}
so in this appendix we will argue that if $R_0E/(GM^2) \gg 1$, $p_{\rm hard} \propto R_0^{-2}$. This dependence clearly comes from the entire phase-space volume, i.e. the normalisation of $f$ in equation \eqref{eqn: outcome distribution}. This is readily seen by observing that for hard binaries, the integration boundaries for $E_b$ are such that $R_0/a_b \gg 1$, so $\theta_{\max}(R_0,E_b,S) = 2\pi$ always, and nothing seemingly depends on $R_0$. But of course, equation \eqref{eqn: outcome distribution} is not normalised, and the normalisation 
\begin{equation}
    N = \iiint \mathrm{d}E_b \mathrm{d}^2 S ~m_b\frac{\theta_{\max}(R_0,E_b,S)\theta_{\max}(R,E_s,\abs{\mathbf{J}-\mathbf{S}}) }{\abs{\mathbf{J}-\mathbf{S}}E_s^{3/2}\abs{E_b}^{3/2}}
\end{equation}
does depend on $R_0$. Obviously, $N=0$ when $R_0 = 0$ and $N = \infty$ when $R_0 = \infty$ -- where there is no cut-off, because the integral over $E_b$ diverges at $E_b = 0$. In fact, we will show that $N \propto R_0^2$ (for a fixed $\kappa$) in the large $R_0$ limit, which will prove that $p_{\rm hard} \propto R_0^{-2}$.

Because $N$ diverges as $R_0 \to \infty$, it is dominated by the region around $E_b = 0$, and in fact by the unbound side, $E_b > 0$ (as we will show below). We therefore consider 
\begin{equation}
\begin{aligned}
    N_\eps & \equiv \int_0^{\eps E}\mathrm{d}E_b \iint \mathrm{d}^2 S \\ & 
    m_b\frac{\theta_{\max}(R_0,E_b,S)\theta_{\max}(R_s,E_s,\abs{\mathbf{J}-\mathbf{S}}) }{\abs{\mathbf{J}-\mathbf{S}}E_s^{3/2}\abs{E_b}^{3/2}}.
\end{aligned}
\end{equation}
We choose $\eps$ such that 
\begin{equation}
    \frac{GM^2}{R_0E} \ll \eps \ll 1,
\end{equation}
whence, $N \approx N_\eps$ to leading order. 
According to \citet[appendix A]{GinatPerets2021a}, the angular momentum integral evaluates to $\scri$, which, in the case $\eps \ll 1$, is $\scri \approx J/2$, and $\theta_{\max}$ becomes $\theta_{ap}$, defined by 
\begin{equation}
    \theta_{ap} = \sqrt{2\frac{R_0}{a_b}+\frac{R_0^2}{a_b^2}} - \arccosh\left(1+\frac{R_0}{a_b}\right).
\end{equation}
Therefore
\begin{equation}
    N_\eps \approx \frac{m_b J}{2}\int_0^{\eps E}\mathrm{d}E_b ~\frac{\theta_{ap}(R_0,E_b)\theta_{ap}(R_s,E_s)}{E_s^{3/2}\abs{E_b}^{3/2}}.
\end{equation}
Again, because $\eps \ll 1$, we have $E_s \approx E$ and $E_b \ll E$, so $\theta_{ap}(R_s,E_s) = \theta_{ap}(R_0,E) \approx 2R_0E/(GM\mu_s)$ (because for $E_b>0$, $R_s = R_0$), so
\begin{equation}
    N_\eps \approx \frac{2m_bJR_0}{GM\mu_sE^2}\int_0^{\eps E}\mathrm{d}E_b ~\frac{\theta_{ap}(R_0,E_b)}{\abs{E_b}^{3/2}}.
\end{equation}

Let us change variables to $x = R_0/a_b = 2R_0E_b/(Gm_b\mu_b)$, and use the integral 
\begin{equation}
\begin{aligned}
    & \int_0^x \frac{1}{y^{3/2}}\left[\sqrt{2y+y^2}-\arccosh(1+y)\right]\mathrm{d}y \\ & 
    = \frac{2 \sqrt{x (x+2)}}{\sqrt{x}}-2 \sqrt{2} \ln \left[2 \sqrt{x}+\sqrt{2} \sqrt{x (x+2)}\right] -4 \sqrt{2}\\ & 
    +\sqrt{2} \ln (x)+2 \sqrt{2} \ln \left(\sqrt{2} \sqrt{x+2}+2\right)+\frac{2 \arccosh(x+1)}{\sqrt{x}},
\end{aligned}
\end{equation}
which grows like $2\sqrt{x}$ at large $x$. These yield
\begin{equation}
    N_\eps \approx \frac{m_bJR_0}{GM\mu_sE^2}\sqrt{\frac{2R_0}{Gm_b\mu_b}}\int_0^{\frac{2R_0\eps E}{Gm_b\mu_b}}\mathrm{d}x ~\frac{\theta_{ap}(x)}{x^{3/2}}.
\end{equation}
Luckily, $\frac{2R_0\eps E}{Gm_b\mu_b} \gg 1$ because of our choice of $\eps$, so we can use the asymptotic $\int_0^x \theta_{ap}(x')x'^{-3/2}\mathrm{d}x' \sim 2\sqrt{x}$; this gives 
\begin{equation}
    N_\eps \approx \frac{m_bJR_0}{GM\mu_sE^{3/2}}\frac{2R_0\sqrt{\eps}}{Gm_b\mu_b} = \frac{4J}{G^2M\mu_s\mu_b}R_0^2\sqrt{\eps}.
\end{equation}
The dependence on $\eps$ is immaterial, but importantly $N_\eps$ -- and therefore $N$ -- grows quadratically with $R_0$ in this limit. 

Now consider the marginally bound binaries, contributing to 
\begin{equation}
\begin{aligned}
    N_{\textrm{mb},\eps} & \equiv \int_{-\eps E}^0\mathrm{d}E_b \iint \mathrm{d}^2 S \\ & 
    m_b\frac{\theta_{\max}(R_0,E_b,S)\theta_{\max}(R_s,E_s,\abs{\mathbf{J}-\mathbf{S}}) }{\abs{\mathbf{J}-\mathbf{S}}E_s^{3/2}\abs{E_b}^{3/2}}.
\end{aligned}
\end{equation}
This integral evaluates similarly, except one difference: while $\theta_{ap}(R_s,E_s) \approx 2R_0E/(GM\mu_s)$ still,  $\theta_{\max}(R_0,E_b,S) \in [0,2\pi]$, because it is the mean anomaly of an elliptic orbit. Consequently, the integral 
\begin{equation}
    \int_0^x \frac{\theta_{\max}(y)}{y^{3/2}}\mathrm{d}y,
\end{equation}
where $y = R_0/a_b$, is now no longer dominated by the upper boundary, so it evaluates to something which depends on $R_0$ weakly, and for $x \gg 1$ it is approximated by
\begin{equation}
    \int_0^{\infty}\frac{\theta_{ap}(y)}{y^{3/2}}\mathrm{d}y = \sqrt{2}(4-\pi) \approx 1.214.
\end{equation}
Therefore, 
\begin{equation}
    N_{\textrm{mb},\eps} \propto R_0^{3/2}
\end{equation}
in this limit. 

Hence, $N \sim N_\eps \propto R_0^2$, and we have the scalings
\begin{align}
    & p_{\rm bin} \sim \frac{N_{\textrm{mb},\eps}}{N_{\eps}} \propto R_0^{-1/2} \\ &
    p_{\rm hard} \propto R_0^{-2}.
\end{align}
These prove the scalings in equation \eqref{eqn:scaling p bin with R0}. 

\begin{figure}
    \centering
    \includegraphics[width=0.49\textwidth]{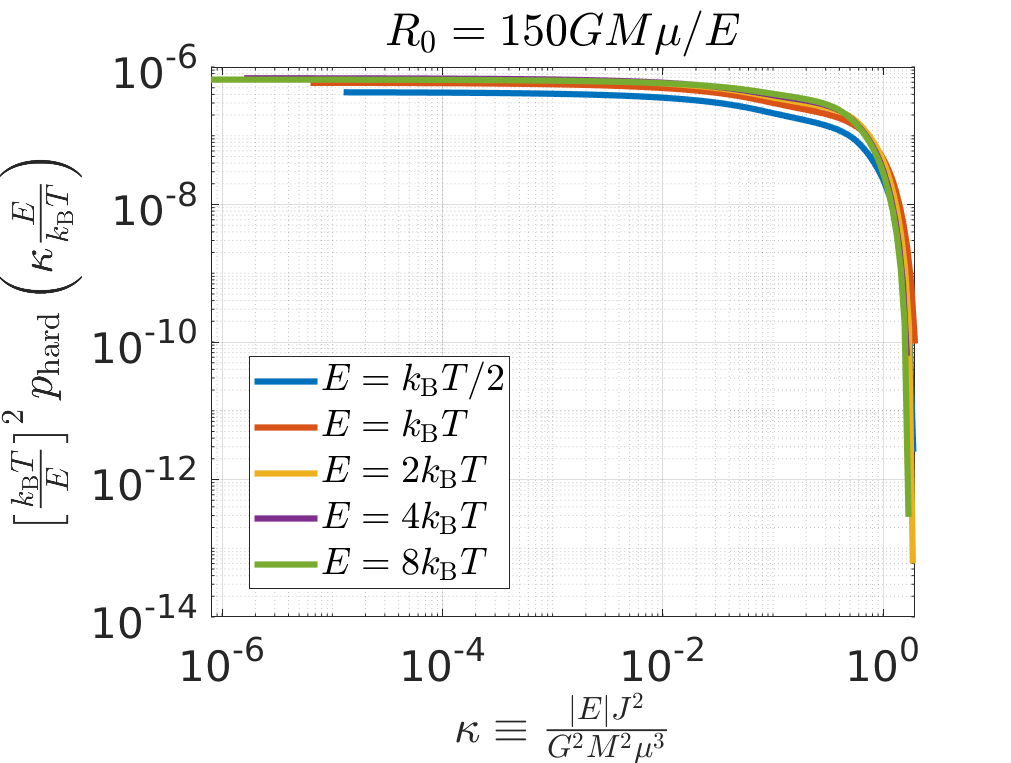}
    \caption{A comparison of the scaling in equation \eqref{eqn:p hard scaling} with a direct integration of equation \eqref{eqn:E_b and S joint distribution} for various values of the hard-soft boundary (i.e. multiples of $k_{\rm B}T$). The scaling is approximately satisfied by $p_{\rm hard}$, with only the highest temperature performing worse; we suspect that this discrepancy is due to the extreme smallness of $p_{\rm hard}$ there.}
    \label{fig:p hard with kBT}
\end{figure}

\begin{figure}
    \centering
    \includegraphics[width=0.49\textwidth]{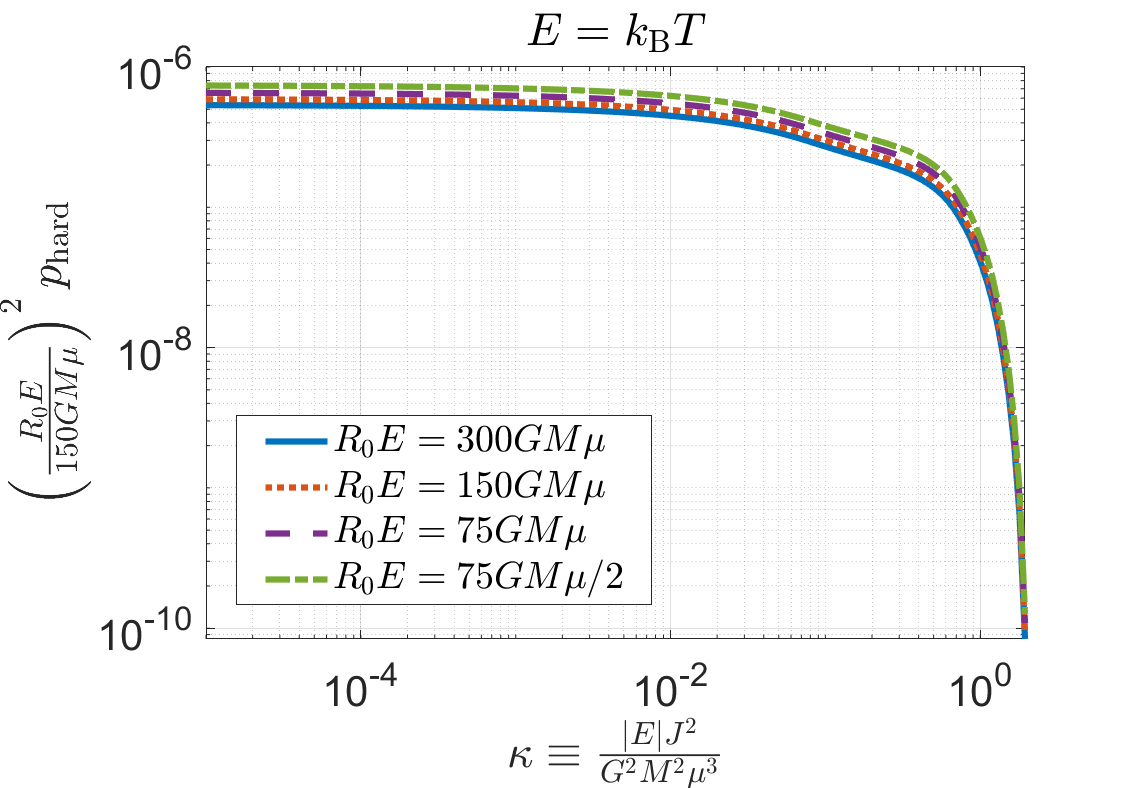}
    \caption{A comparison of the scaling in equation \eqref{eqn:p hard scaling} with a direct integration of equation \eqref{eqn:E_b and S joint distribution} for various values of the cut-off $R_0$.}
    \label{fig:p hard with R0}
\end{figure}

Having derived these scaling relations analytically, let us test them by comparing them with direct numerical integration of equation \eqref{eqn: outcome distribution} or equation \eqref{eqn:E_b and S joint distribution}, each time for different values of $R_0$ and $E/(k_{\rm B}T)$. Figure \ref{fig:p hard with kBT} shows the scaling with temperature, while figure \ref{fig:p hard with R0} shows the scaling with $R_0$. Both agree qualitatively with the analytical scalings. 

\begin{figure}
    \centering
    \includegraphics[width=0.49\textwidth]{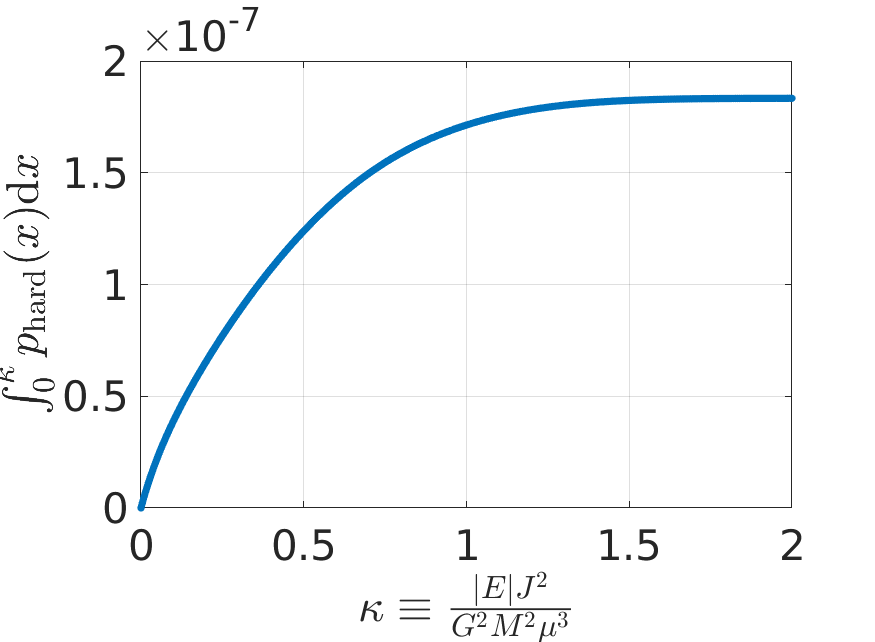}
    \caption{A plot of the integral of $p_{\rm hard}$ over $\kappa$, for $R_0 E = 150 GM\mu$. See text for details.}
    \label{fig:integrated p hard dkappa}
\end{figure}

Before moving on to calculate the rate, the only remaining function we have to compute for the rate is $P_{\rm hard}(\kappa) \equiv \int_0^\kappa p_0(\kappa';d_0)\mathrm{d}\kappa'$ (see equation \eqref{eqn: total p hard} below). We do so for the same value of $R_0$ as in the top panel of figure \ref{fig:binary formation probability}. The result is shown in figure \ref{fig:integrated p hard dkappa}. To simplify calculations, we tabulate this integral and interpolate between its values, whenever it is required in the rate calculation. 

\section{Rate Calculation}
\label{appendix: rate calculation}
The purpose of this appendix is the calculation of the binary formation probability in equation \eqref{eqn: probability full}, and likewise the similar calculation for the hard binary formation probability. While doing so, we will derive the scalings \eqref{eqn:scaling laws for the full probability}.

\subsection{Hard Binaries}
Let us start with $P_{\rm hard}$: we first change variables from $\mathbf{r}$, $\mathbf{v}$, $\mathbf{R}$, $\mathbf{V}$ to hyperbolic Delaunay variables \citep{Floria1995},\footnote{Here we denote what that paper called $l$ and $L$ by $\theta_c$ and $J_c$. The super-script (or sub-script) $i$ is used to refer to the initial values, and the two other action variables are the absolute value and the $\zhat$ component of the angular momentum.} which are treated here as mere phase-space co-ordinates -- we do not claim that the orbit is in fact two hyperbolas, but just use these as a way to sample the initial conditions. The Jacobian 
is 
\begin{align}
    & \mathrm{d}^3V \mathrm{d}^3R = \mu_s^{-3}\mathrm{d}^3 \theta_s \mathrm{d}^3 J_s \\ & 
    \mathrm{d}^3v \mathrm{d}^3r = \mu_b^{-3}\mathrm{d}^3 \theta_b \mathrm{d}^3 J_b.
\end{align}
Moreover, the angle integral is immediately carried out to yield $(2\pi)^4 \times (2\theta_{\max}^b)(2\theta_{\max}^s)$, where $\theta_{\max}$ incorporates the cut-offs in \eqref{eqn:cut-offs initial}; the factors of $2$ come from the possibility of negative mean anomalies, as well as positive ones. We can approximate as above (cf. appendix A of \citealt{GinatPerets2021a})
\begin{equation}
    \theta_{\max} \approx \theta_{ap}\xi(b),
\end{equation}
with 
\begin{equation}
    \theta_{ap}(x) = \sqrt{2x+x^2} - \arccosh(1+x),
\end{equation}
and $b \equiv S/S_{\max}$ [resp. $L/L_{\max}$], while
\begin{equation}
    \xi(b) \equiv \sqrt{1-b^2}(1+2b^2),
\end{equation}
and $x = R_{\max}/a$, where $a$ is the semi-major axis and $R_{\max}$ is the appropriate upper bound on the pericentre distance. 

Equation \eqref{eqn: probability full}, upon replacing $P_{\rm bin}$ with $P_{\rm hard}$, becomes
\begin{equation}
\begin{aligned}
    P_{\rm hard} & = \frac{8\pi \mu_b^{-3}\mu_s^{-3}}{V_0^2}\left(\frac{m_1m_2m_3}{M}\right)^{3/2}\frac{1}{(k_{\rm B}T)^3} \\ & 
    \times \int \mathrm{d}J_{c,s}^i\int \mathrm{d}J_{c,b}^i \int \mathrm{d}^2S_i \int \mathrm{d}^2L_i ~ \xi(b_b)\xi(b_s) \theta_{ap}^b\theta_{ap}^s \\ & 
    \times p_{\rm hard} \mathrm{e}^{-\frac{E}{k_{\rm B}T}},
\end{aligned}
\end{equation}
where $J_c^i$ is the Delaunay action conjugate to the mean anomaly, and $\mathbf{S}_i$ and $\mathbf{L}_i$ are the initial angular momenta. The $S_i$ integral simply evaluates to \begin{equation}
    2\mu_b R_0^2 E_{b,i}\int_0^12b\xi(b)\mathrm{d}b = 3.6\mu_b R_0^2 E_{b,i}
\end{equation}

We can now change variables from an integral over $\mathbf{L}_i$ to an integral over the total angular momentum, $\mathbf{J}$, which goes from $0$ to $J_{\max}$, which is nothing but (by definition of $\kappa$)
\begin{equation}
    J_{\max} = \frac{GM\mu^{3/2}\sqrt{\kappa_{\max}}}{\sqrt{E}}.
\end{equation}
Luckily, for equal masses, this, in conjunction with equation \eqref{eqn:cut-offs initial}, implies that $b_s$ is small, whence $\xi(b_s) \approx 1$.
Hence, 
\begin{equation}
\begin{aligned}
    P_{\rm hard} & = \frac{20\pi \mu_b^{-3}\mu_s^{-3}}{V_0^2}\left(\frac{m_1m_2m_3}{M}\right)^{3/2}\frac{\mu_bR_0^2}{(k_{\rm B}T)^3} \\ & 
    \times \int \mathrm{d}J_{c,s}^i\int \mathrm{d}J_{c,b}^i  \int_0^{J_{\max}} \mathrm{d}^2J ~ E_b \theta_{ap}^b\theta_{ap}^s p_{\rm hard} \mathrm{e}^{-\frac{E}{k_{\rm B}T}}.
\end{aligned}
\end{equation}
We will now change variables from $J_{c,s}^i$ and $J_{c,b}^i$ to the initial energies $E_{s,i}$ and $E_{b,i}$, and from $J$ to $\kappa$; the Jacobians for this transformation are
\begin{align}
    & \mathrm{d}\kappa = \frac{2EJ}{G^2M^2\mu_s^3}\mathrm{d}J \\ & 
    \mathrm{d}J_{c,s}^i = -\frac{GM\mu_s^{3/2}}{\sqrt{8}E_{s,i}^{3/2}}\mathrm{d}E_{s,i} \\ &
    \mathrm{d}J_{c,b}^i = -\frac{Gm_b\mu_b^{3/2}}{\sqrt{8}E_{b,i}^{3/2}}\mathrm{d}E_{b,i}.
\end{align}
We will also henceforth drop the $i$ where it isn't confusing to do so. This transforms $P_{\rm hard}$ into
\begin{equation}
\begin{aligned}
    P_{\rm hard} & = \frac{5\pi \mu_b R_0^2G^4M^3m_b\mu_s^3}{2V_0^2(k_{\rm B}T)^3}\left(\frac{m_1m_2m_3}{M\mu_s\mu_b}\right)^{3/2}\\ & 
    \times \iint_0^{\infty}\frac{\mathrm{d}E_b\mathrm{d}E_s}{(E_bE_s)^{3/2}} \int_0^{\kappa_{\max}} \mathrm{d}\kappa ~ \frac{E_{b}}{E} \theta_{ap}^b\theta_{ap}^s p_{\rm hard} \mathrm{e}^{-\frac{E}{k_{\rm B}T}}.
\end{aligned}
\end{equation}
Finally, the scaling in equation \eqref{eqn:p hard scaling} implies that 
\begin{equation}\label{eqn: total p hard}
\begin{aligned}
    P_{\rm hard} & = \frac{5\pi \mu_b d_0^2G^4M^3m_b\mu_s^3}{2V_0^2(k_{\rm B}T)^4}\left(\frac{m_1m_2m_3}{M\mu_s\mu_b}\right)^{3/2}\\ & 
    \times \iint_0^{\infty}\frac{\mathrm{d}E_b\mathrm{d}E_s}{(E_bE_s)^{3/2}} ~ E_{b} \theta_{ap}^b\theta_{ap}^s \mathrm{e}^{-\frac{E}{k_{\rm B}T}} \\\ & 
    \times \int_0^{\kappa_{\max}\frac{k_{\rm B}T}{E}}\mathrm{d}u ~p_{0}(u;d_0) .
\end{aligned}
\end{equation}
Equation \eqref{eqn: total p hard} is the equation that we integrate numerically for the plot in figure \ref{fig:probability}, specifically, the line denoted by $P_{\rm hard,1}$ there. Let us now show that this yields a $\zeta ^5 \propto \chi_1^{-5}$ scaling.

\subsubsection{Scaling at $\zeta \to 0$}
First, note that $V_0^{-2} \propto \zeta^6$. In the limit of small $\zeta$, i.e. large $R_0$, the only other dependence on $R_0$ in equation \eqref{eqn: total p hard} can come from $\theta_{ap}^b\theta_{ap}^s$. These are obtained from the constraints in equation \eqref{eqn:cut-offs initial}, which imply that in fact $\theta_{ap}^s$ is independent of $R_0$, while the argument of $\theta_{ap}^b$, namely 
\begin{equation}
    \frac{2R_0E_b}{Gm_b\mu_b} \gg 1
\end{equation}
is very large. Consequently, 
\begin{equation}
    \theta_{ap}^b \sim \frac{2R_0E_b}{Gm_b\mu_b}
\end{equation}
in this limit (recall that this is the Delaunay angle of a hyperbolic trajectory, which is unbounded).
Together, we find
\begin{equation}
    \begin{aligned}
        P_{\rm hard,1} & \sim R_0^{-5} \times \frac{45 d_0^2G^3M^3\mu_s^3}{16\pi(k_{\rm B}T)^4}\left(\frac{m_1m_2m_3}{M\mu_s\mu_b}\right)^{3/2}\\ & 
        \times \iint_0^{\infty}\frac{\mathrm{d}E_b\mathrm{d}E_s}{E_s^{3/2}} ~ E_{b}^{1/2} \theta_{ap}^s \mathrm{e}^{-\frac{E}{k_{\rm B}T}} \\ & 
        \times \int_0^{\kappa_{\max}\frac{k_{\rm B}T}{E}}\mathrm{d}u ~p_{0}(u;d_0).
    \end{aligned}
\end{equation}

Observe, that if we had used the second condition in equation \eqref{eqn:cut-offs initial}, instead of the first, we would have had $\theta_{ap}^s \sim \sqrt{r}$, which is dominated by the large $r$ r\'{e}gime, i.e. $\theta_{ap}^s \sim \sqrt{R_0}$. This extra power of $1/2$ would have resulted in an overall $P_{\rm hard} \propto R_0^{-9/2}$. This is what is plotted in figure \ref{fig:probability} as $P_{\rm hard,2}$, \emph{viz.}
\begin{equation}\label{eqn: total p hard 2}
\begin{aligned}
    P_{\rm hard,2} & = \frac{5\pi \mu_b d_0^2G^4M^3m_b\mu_s^3}{2V_0^2(k_{\rm B}T)^4}\left(\frac{m_1m_2m_3}{M\mu_s\mu_b}\right)^{3/2}\\ & 
    \times \iint_0^{\infty}\frac{\mathrm{d}E_b\mathrm{d}E_s}{(E_bE_s)^{3/2}} ~ E_{b} \theta_{ap}^b\theta_{ap,\textrm{soft}}^s \mathrm{e}^{-\frac{E}{k_{\rm B}T}} \\\ & 
    \times \int_0^{\kappa_{\max}\frac{k_{\rm B}T}{E}}\mathrm{d}u ~p_{0}(u;d_0),
\end{aligned}
\end{equation}
with $\theta_{ap,\textrm{soft}}^s$ has as its argument 
\begin{equation}
    \frac{1}{a_b}\frac{2Gm_s\sqrt{\mu_s}}{\sqrt{2E_s}}.\sqrt{\frac{r}{2Gm_b}}.
\end{equation}

We also remark that this is the scaling reported by \cite{Atallahetal2024}, but that this scaling would imply that the rate does not converge to a constant at arbitrarily large impact parameters. For this reason, we use $P_{\rm hard,1}$ as $P_{\rm hard}$ in the rest of this paper. 

\subsection{Soft Binaries}
We repeat the same procedure as above but only change variables to Delaunay variables in the `outer' orbit, denoted by $s$. This results in 
\begin{equation}
    \begin{aligned}
        P_{\rm bin} & = \frac{1}{\pi V_0^2 (k_{\rm B}T)^3} \left(\frac{m_1m_2m_3}{M}\right)^{3/2} \\ & 
        \times \int \mathrm{d}^3r\mathrm{d}^3v \int \mathrm{d}J_{c,s}^i \theta_{ap}^s \int \mathrm{d}^2 L_i p_{\rm bin} \mathrm{e}^{-\frac{E}{k_{\rm B}T}}.
    \end{aligned}
\end{equation}
The integral over $\mathbf{L}_i$ is just $R_{\max}^2V^2\mu_s^2$, where 
\begin{equation}
    R_{\max} \equiv \frac{2Gm_s}{V\sqrt{2Gm_b/r}}.
\end{equation}
Additionally, by the arguments in \S \ref{subsec: hard binary formation} and appendix \ref{appendix: p hard and R0}, $p_{\rm bin}$ scales like $\sqrt{d_0/R_0}$. Besides, the $\kappa$ dependence is very weak, as the inequality $\kappa \leq \kappa_{\max}$ is satisfied for all energies (for equal masses), if $\zeta < 20.25$; we are concerned with $\zeta \ll 1$, so we just take $p_{\rm bin}(\kappa) \approx p_{\rm bin}(\kappa = 0)$. 

Repeating the above procedure yields eventually
\begin{equation}\label{eqn: total p bin}
    \begin{aligned}
        P_{\rm bin} & = \frac{2\sqrt{8}\pi GM\mu_s^{7/2}}{V_0^2(k_{\rm B}T)^3}  \left(\frac{m_1m_2m_3}{M\mu_s^2}\right)^{3/2} \sqrt{\frac{d_0}{R_0}}\overline{p}_{\rm bin}(d_0) \\ & 
        \times \int_0^\infty \mathrm{d}v\int_0^{R_0}\mathrm{d}r \int_0^\infty \mathrm{d}E_s^i ~\frac{v^2r^2}{(E_s^i)^{3/2}} \theta_{ap}^s V^2R_{\max}^2 \mathrm{e}^{-\frac{E}{k_{\rm B}T}}.
    \end{aligned}
\end{equation}
This equation is used in figure \ref{fig:probability} to give the line denoted by $P_{\rm bin}$. 

Let us now show that it scales like $\zeta^2$ at $\zeta \ll 1$: from \eqref{eqn:cut-offs initial}, $R_{\max}^2 \propto r$, and $\theta_{ap}^s$ goes like $\sqrt{r}$. The integral over $r$ is then proportional to $R_0^{9/2}$. Together with the $R_0^{-13/2}$ already present, this becomes $R_0^{-2}$. This scaling agrees with that obtained originally by \cite{AarsethHeggie1976} analytically and numerically, and more recently in the simulations of \cite{Atallahetal2024}.

\bsp	
\label{lastpage}
\end{document}